\documentclass[12pt]{article}

\usepackage{amsmath}
\usepackage{amssymb}
\usepackage{latexsym}
\usepackage{graphicx}
\usepackage{epsfig}

\def\beq{\begin{equation}}
\def\eeq{\end{equation}}
\def\bea{\begin{eqnarray}}
\def\eea{\end{eqnarray}}

\begin{document}

\begin{titlepage}

\vspace*{1cm}
\begin{center}
{\bf \Large Emission of Massive Scalar Fields\\[2mm] by a
Higher-Dimensional Rotating Black-Hole}

\bigskip \bigskip \medskip

{\bf P. Kanti} and
{\bf N. Pappas}

\bigskip
{\it Division of Theoretical Physics, Department of Physics,\\
University of Ioannina, Ioannina GR-45110, Greece}

\bigskip \medskip
{\bf Abstract}
\end{center}
We perform a comprehensive study of the emission of massive scalar
fields by a higher-dimensional, simply rotating black hole both in
the bulk and on the brane. We derive approximate, analytic results
as well as exact numerical ones for the absorption probability, and
demonstrate that the two sets agree very well in the low and
intermediate-energy regime for scalar fields with mass $m_\Phi \leq$
1 TeV in the bulk and $m_\Phi \leq$ 0.5 TeV on the brane. The
numerical values of the absorption probability are then used to derive
the Hawking radiation power emission spectra in terms of the number
of extra dimensions, angular-momentum of the black hole and mass
of the emitted field. We compute the total emissivities in the bulk
and on the brane, and demonstrate that, although the brane channel
remains the dominant one, the bulk-over-brane energy ratio is
considerably increased (up to 33\%) when the mass of the emitted
field is taken into account.

\end{titlepage}


\section{Introduction}

The postulation of the existence of additional spacelike dimensions
in nature, that can be as large as a few micrometers \cite{ADD}
or even infinite in size \cite{RS}, has led to the idea of a
higher-dimensional gravitational theory with a fundamental energy
scale $M_*$ much smaller than the traditional Planck scale $M_P$.
If this can be realised with $M_*$ close to the TeV scale, present
of future experiments may accelerate particles at energies beyond
this new gravity scale. This will unavoidably lead to the
occurence of strong gravity effects in particle collisions and
the production of heavy final states, including miniature black holes
\cite{creation}.

The lifetime of these black holes is expected to be very short as they
instantaneously decay via the emission of Hawking radiation \cite{Hawking}
(for detailed reviews of their properties, see \cite{Kanti, reviews}).
Since these black holes will be created and decay in front of our detectors,
it is anticipated that the emission of Hawking radiation will be the
main obervable signature of their creation and, at the same time, a
manifestation of the existence of additional spacelike dimensions in nature.
As a result, the study of the emission of radiation by higher-dimensional
black holes has been the subject of an intensive research activity over
the last years. This includes the emission from both spherically-symmetric
\cite{KMR, HK1, Barrau, Jung, BGK, Naylor, Park, Cardoso, CEKT1, Dai} and
rotating \cite{FS-rot, IOP, Nomura, HK2, IOP2, Jung-super, Jung-rot, DHKW,
CKW, CDKW, CEKT2, CEKT3, CEKT4, CDKW2, KKKPZ, CDKW3} black holes in the form of
zero and non-zero spin fields.

In order to simplify the analysis, the emitted fields are assumed to
be minimally-coupled to gravity but otherwise free as well as massless.
Nevertheless, in the context of the four-dimensional analysis \cite{Page}
it was found that for certain particles and mass of the black hole, the
particle mass can significantly (up to 50\%) suppress the emission rate.
Recently, a set of works \cite{Sampaio} has addressed the question of
the role of the mass of the emitted field (as well as that of the charge)
for emission on the brane by a higher-dimensional black hole. Here, we
extend this analysis by considering the case of a higher-dimensional
black hole with a non-vanishing angular momentum emitting massive scalar
fields. We perform a
comprehensive study of the absorption probability and energy emission
rate for a range of values of the mass of the emitted field, number
of extra dimensions, and angular momentum of the black hole. By
integrating over the entire frequency range, we compute the total
emissivities and obtain the suppression factors in each case. We
also consider the cases of both bulk and brane emission, and pose
the additional question of whether the presence of the mass of the
emitted field can affect the bulk-over-brane energy ratio and threaten
the dominance of the brane channel.

The outline of this paper is as follows: In section 2, we study the
emission of massive scalar fields by a higher-dimensional, simply rotating
black hole in the bulk; we compute the value of the absorption probability
both analytically and numerically and compare the two sets of results;
finally, we derive the exact energy emission spectra and discuss their
behaviour. In section 3, we turn to the brane and perform the same tasks.
The total emissivities for bulk and brane emission are derived in section
4, and the bulk-over-brane ratio is computed for a large number of values
of the parameters of the theory. We close with our conclusions in section 5.


\section{Emission of Massive Scalars in the Bulk}


In this work, we will consider the case of a higher-dimensional, neutral,
simply rotating black hole whose gravitational background is described
by the following form of the Myers-Perry solution \cite{MP}
\begin{eqnarray}
&~& \hspace*{-3cm}ds^2 =
-\biggl(1-\frac{\mu}{\Sigma\,r^{n-1}}\biggr) dt^2 - \frac{2 a \mu
\sin^2\theta}{\Sigma\,r^{n-1}}\,dt \, d\varphi
+\frac{\Sigma}{\Delta}\,dr^2 +\Sigma\,d\theta^2 \nonumber \\[2mm]
\hspace*{2cm} &+& \biggl(r^2+a^2+\frac{a^2 \mu
\sin^2\theta}{\Sigma\,r^{n-1}}\biggr) \sin^2\theta\,d\varphi^2 +
r^2 \cos^2\theta\, d\Omega_{n}^2, \label{rot-metric-bulk}
\end{eqnarray}
where
\begin{equation}
\Delta = r^2 + a^2 -\frac{\mu}{r^{n-1}}\,, \qquad \Sigma=r^2
+a^2\,\cos^2\theta\,, \label{Delta}
\end{equation}
and $d\Omega_n^2(\theta_1,\theta_2,\ldots,\theta_{n-1},\phi)$ is
the line-element on a unit $n$-sphere. The above line-element is
expected to describe black holes created by an on-brane collision of
particles that acquire only one non-zero angular momentum component,
parallel to our brane. The black hole's mass $M_{BH}$ and angular
momentum $J$ are then related to the parameters $\mu$ and $a$,
respectively, as follows
\begin{equation}
M_{BH}=\frac{(n+2) A_{n+2}}{16 \pi G_D}\,\mu\,,  \qquad
J=\frac{2}{n+2}\,M_{BH}\,a\,, \label{def}
\end{equation}
with $G_D$ being the $(4+n)$-dimensional Newton's constant, and
$A_{n+2}$ the area of a $(n+2)$-dimensional unit sphere given by
$A_{n+2}=2 \pi^{(n+3)/2}/\Gamma[(n+3)/2]$. The black hole's horizon
radius $r_h$ follows from the equation $\Delta(r_h)=0$, and may
be written as $r_h^{n+1}=\mu/(1+a_*^2)$, where $a_*=a/r_h$.

A massive scalar field, with mass $m_\Phi$, propagating in the gravitational
background (\ref{rot-metric-bulk}) will satisfy the equation of motion
\beq \frac{1}{\sqrt{-G}}\,\partial_M\left(\sqrt{-G}\,G^{MN}\partial_N \Phi\right)
-m_\Phi^2 \Phi=0\,, \label{field-eq-bulk}\eeq
where $G_{MN}$ the higher-dimensional metric tensor and $G$ its
determinant satisfying the relation
\beq \sqrt{-G}=\Sigma \sin\theta \, r^{n}\cos^{n}\theta
\prod_{i=1}^{n-1} \sin^{i}\theta_i\,. \eeq
Even in the presence of the mass term, the above equation can be separated
\cite{FS-rot, FS} by assuming the factorised ansatz
\beq \Phi = e^{-i\omega
t}e^{im\varphi}R(r)\,S(\theta)\,Y_{ln}(\theta_1,\ldots,\theta_{n-1},\phi)
\,, \eeq
where $Y_{ln}(\theta_1,\ldots,\theta_{n-1},\phi)$ are the
hyperspherical harmonics on the $n$-sphere that satisfy the
equation \cite{Muller,IMU}
\begin{equation}
\sum_{k=1}^{n-1} \frac{1}{\prod_{i=1}^{n-1}
\sin^i\theta_i}\,\partial_{\theta_k} \left[\left(\prod_{i=1}^{n-1}
\sin^i\theta_i\right)\,\frac{\partial_{\theta_k} Y_{ln}}
{\prod_{i>k}^{n-1} \sin^2\theta_i}\right] +
\frac{\partial_{\phi\phi} Y_{ln}} {\prod_{i=1}^{n-1}
\sin^2\theta_i}+l(l+n-1)\,Y_{ln}=0\,. \label{hyper-eq}
\end{equation}
The functions $R(r)$ and $S(\theta)$ in turn satisfy the following decoupled
radial and angular equation
\begin{equation}
\frac{1}{r^n}\,\partial_r \left(r^n\Delta\,\partial_r R\right) +
\left(\frac{K^2}{\Delta} - \frac{l(l+n-1)a^2}{r^2} -
\tilde\Lambda_{j\ell m} - m_\Phi^2r^2 \right)R = 0 \,,\label{eq:radial-bulk}
\end{equation}
\begin{equation}
\frac{1}{\sin\theta \cos^n
\theta}\,\partial_\theta\left(\sin\theta \cos^n \theta
\partial_\theta S\right) +\left(\tilde\omega^2 a^2 \cos^2\theta - \frac{m^2}{\sin^2\theta}
- \frac{l(l+n-1)}{\cos^2\theta} + \tilde E_{j\ell m}\right)S = 0
\,, \label{eq:hdspheroidal}
\end{equation}
respectively. In the above,
\begin{equation}
K=(r^2+a^2)\,\omega-am\,,  \qquad \tilde\Lambda_{j\ell m}=\tilde
E_{j\ell m}+a^2\omega^2-2am\omega\,. \label{eq:K}
\end{equation}
For the above decoupling to take place, the angular function $S(\theta)$
needs to satisfy a modified higher-dimensional spheroidal harmonics
equation: compared to the massless case \cite{Eigenvalues1}, it has
the energy $\omega$ replaced by the momentum
$\tilde\omega \equiv \sqrt{\omega^2-m_\Phi^2}$. Then, the massive
angular eigenvalue $\tilde E_{j\ell m} (a\tilde\omega)$ is related
to the massless one $E_{j\ell m} (a \omega)$ by merely a shift of its
argument: $a \omega \rightarrow a\tilde\omega$. Here, we will employ
the power-series expression of the angular eigenvalue \cite{Eigenvalues1}
in terms of the parameter $(a \omega)$ which, under the aforementioned shift
and up to 5th order, takes the form
\bea
\tilde E_{jlm} &=& j\,(j+n+1)- (a\tilde\omega)^2\,\frac{[-1 + 2 l (l - 1)
+ 2 j (j + 1) - 2 m^2 + 2 n (j + l) + n^2]}{(2 j + n - 1)\,(2 j + n + 3)}
\nonumber \\[1mm] &+&
(a\tilde\omega)^4\,\left\{\frac{(l - j + |m|) (l + j - |m| + n - 1)}
{16 (2j + n - 3) (2 j + n - 1)^2}\biggl[(2 + l - j + |m|) (l + j -|m|+ n - 3)
\right. \nonumber\\[1mm]
&-& \left. 4 (2 j + n - 3)\frac{[-1 + 2 l (l - 1) + 2 j (j + 1) - 2 m^2 +
 2 n (j + l) + n^2]}{(2 j + n - 1) (2 j + n + 3)}\biggr]\right. \nonumber\\[1mm]
&-& \left.  \frac{(l - j + |m| - 2) (l + j + n - |m| + 1)}{16 (2 j + 5 + n)
(2 j + n + 3)^2}\biggl[(l - j + |m| - 4) (j + l + n - |m| + 3) \right. \nonumber\\[1mm]
&+& \left. 4 (2 j + n + 5)\frac{[-1 + 2 l (l - 1) + 2 j (j + 1) - 2 m^2 +
 2 n (j + l) + n^2]}{(2 j + n - 1) (2 j + n + 3)}\biggr]\right\} \nonumber\\[1mm]
&+& {\cal O}\Bigl((a \tilde \omega)^6\Bigr)\,. \label{Eigen-bulk}
\eea
The analytic form of the angular eigenvalue was studied in detail in
the context of previous works focusing on the emission of massless scalars
\cite{CEKT4, CDKW2} and gravitons \cite{KKKPZ} in the bulk. It was found
that its value, when terms up to 5th order or higher are kept, is remarkably
close to the exact numerical value and that considerable deviations appear
only for a very large angular momentum of the black hole or energy of the
emitted particle, that lie beyond the range of values considered in this work.
For this reason, the analytic form (\ref{Eigen-bulk}) of the
angular eigenvalue will be employed in the derivation of the absorption
probability in both an analytic and numerical method. We should still
demand of course the convergence of the power series by imposing restrictions
on the allowed values of the integer parameters $(j,\ell,m)$ that specify
the emission mode\,: $m$, that denotes the angular momentum of the mode
along our brane, may take any integer value while $\ell$ and $j$ -- the
angular momentum number in the $n$-sphere and total angular
momentum number, respectively -- may take any positive or zero integer value
provided \cite{Eigenvalues1} that $j\geq \ell+|m|$ and $\frac{j-(\ell+|m|)}{2}
\in \{0,\mathbb{Z}^+\}$.


\subsection{The Absorption Probability in the Bulk}

For the derivation of the absorption probability $|{\cal A}_{j \ell m}|^2$
we need the solution for the radial function $R(r)$. We will first solve
Eq. (\ref{eq:radial-bulk}) analytically by using an approximate method,
and we will derive an analytic expression for the absorption probability
which in principle is valid in the low-energy and low-angular-momentum limit.
We will then solve the same equation numerically to derive the exact value
of $|{\cal A}_{j \ell m}|^2$, that will subsequently be used to derive the
Hawking radiation spectrum. The two sets of results will be compared, and the
validity of the approximate method will be studied in terms of the
value of the angular-momentum parameter $a$, number of extra dimensions
$n$ and mass of the emitted particle $m_\Phi$.

The approximate analytic method amounts to solving the radial equation
in the two asymptotic regimes, those of the black-hole horizon and
far away from it, and then matching them in an intermediate regime.
Apart from the appearance of the mass parameter $m_\Phi$, the analysis
is very similar to the one for the emission of massless scalar fields
in the bulk which has already appeared in the literature \cite{CEKT4}.
Therefore, here we briefly present the analysis and results giving
emphasis to the differences arising due to the presence of the mass term.

In terms of the new radial variable $r \rightarrow f(r) = \Delta(r)/(r^2+a^2)$
\cite{CEKT2, CEKT3}, the radial equation (\ref{eq:radial-bulk}) near the
horizon ($ r \simeq r_h$) takes the form
\begin{equation}
f\,(1-f)\,\frac{d^2R}{df^2} + (1-D_*\,f)\,\frac{d R}{df} +
\biggl[\,\frac{K^2_*}{A_*^2\,f (1-f)}-\frac{C_*}{A_*^2\,(1-f)}\,\biggr] R=0\,,
\label{eq:NH-1-bulk}
\end{equation}
where $A_* \equiv (n+1) + (n-1)a_*^2$, $D_* \equiv 1 - 4 a_*^2/A_*^2$, while
$K_*$ and $C_*$ are defined as
\begin{equation}
K_* = (1+a_*^2) \omega r_h - a_* m\,, \qquad C_*=\Bigl[\ell(\ell+n-1)a_*^2 +
\tilde\Lambda_{j\ell m} + m_\Phi^2r_{h}^2\Bigr]\,(1+a_*^2)\,, \label{eq:Kstar}
\end{equation}
respectively. By employing the transformation $R_{NH}(f)=f^{\alpha}\,(1-f)^\beta\,F(f)$
the above equation takes the form of a hypergeometric differential equation
\cite{Abramowitz} as long as
\begin{equation}
\alpha_\pm = \pm \frac{i K_*}{A_*}\,, \qquad
 \beta  = \frac{1}{2}\,\biggl[\,(2-D_*) -
\sqrt{(D_*-2)^2 - 4\left(\frac{K_*^2 - C_*}{A_*^2}\right)}\,\,\biggr]. \label{beta-bulk}
\end{equation}
The radial function $R_{NH}(f)$ must satisfy the boundary condition that no
outgoing modes exist near the black-hole horizon which then reduces the general
solution of the hypergeometric equation to the physically acceptable one
\beq R_{NH}(f)=A_-\,f^{\alpha}\,(1-f)^\beta\,F(a,b,c;f)\,,
\label{NH-bulk} \eeq
with $a=\alpha + \beta +D_*-1$, $b=\alpha + \beta$, $c=1 + 2 \alpha$
and $A_-$ an integration constant. Indeed, we may easily check that in the
limit $r\rightarrow r_h$ (or equivalently $f \rightarrow 0$), and
by making the choice $\alpha=\alpha_-$, we obtain
\beq R_{NH}(f) \simeq A_-\,f^{- i K_*/A_*} = A_-\,e^{- i ky}\,.
\label{near-asym}
\end{equation}
that has a form of an incoming plane-wave, as expected, in terms of a tortoise-like
coordinate defined by $y=r_h (1+a_*^2)\ln(f)/A_*$. In the above, $k$ is given by
\beq k \equiv \frac{K_*}{r_h (1+a_*^2)}=\omega - m \Omega_h =\omega - \frac{m a}{r_h^2
+a^2}\,, \label{k} \eeq
where $\Omega_h$ is the rotation velocity of the black hole.

In the far-field regime ($r \gg r_h$), the substitution $R(r) =
r^{-\left(\frac{n+1}{2}\right)} \tilde R(r)$ brings Eq.
(\ref{eq:radial-bulk}) into the form of a Bessel equation
\cite{Abramowitz}
\beq \frac{d^2\tilde R}{dz^2}+ \frac{1}{z}\frac{d \tilde
R}{dz}+\left(1 - \frac{\tilde E_{j\ell m}+a^2\tilde\omega^2+
\left(\frac{n+1}{2}\right)^2}{z^2} \right)\tilde R=0 \,,
\label{FF-eq} \eeq
in terms of $z \equiv \tilde\omega r$, with solution
\beq
R_{FF}(r)=\frac{B_1}{r^{\frac{n+1}{2}}}\,J_{\nu}\,(\tilde\omega r)
+ \frac{B_2}{r^{\frac{n+1}{2}}}\,Y_{\nu}\,(\tilde\omega r) \,.
\label{FF-bulk} \eeq
In the above, $J_\nu$ and $Y_\nu$ are the Bessel functions of the first
and second kind, respectively, and $\nu=\sqrt{\tilde E_{j\ell
m}+a^2\tilde\omega^2+\left(\frac{n+1}{2}\right)^2}$.

We now need to smoothly match the two asymptotic solutions
(\ref{NH-bulk}) and (\ref{FF-bulk}) in an intermediate regime. The
near-horizon solution (\ref{NH-bulk}) must first be shifted, so that
its argument changes from $f$ to $(1-f)$, and subsequently expanded in the
$r \gg r_h$ limit. Then, it takes the polynomial form
\beq R_{NH}(r) \simeq A_1\, r^{\,-(n+1)\,\beta} +
A_2\,r^{\,(n+1)\,(\beta + D_*-2)}\,, \label{NH-stretched} \eeq
with $A_1$ and $A_2$ defined as \bea A_1 &=& A_-
\left[(1+a_*^2)\,r_h^{n+1}\right]^\beta \,
\frac{\Gamma(c)\Gamma(c-a-b)}{\Gamma(c-a)\Gamma(c-b)}\,, \nonumber\\[1mm]
A_2 &=& A_- \left[(1+a_*^2)\,r_h^{n+1}\right]^{-(\beta + D_*-2)}
\, \frac{\Gamma(c)\Gamma(a+b-c)}{\Gamma(a)\Gamma(b)}\,. \eea
The far-field solution (\ref{FF-bulk}) is in turn expanded to small values
of $r$ leading to
\beq R_{FF}(r) \simeq \frac{B_1\left(\frac{\tilde\omega r}{2}
\right)^\nu}{r^{\frac{n+1}{2}} \,\Gamma(\nu+1)}- \frac{B_2}{\pi \,
r^{\frac{n+1}{2}}}\,\frac{\Gamma(\nu)}{\left(\frac{\tilde\omega
r}{2} \right)^{\nu}} \,. \label{FF-stretched} \eeq
The two polynomial forms match perfectly if we take the small $a_*$ and
$\tilde\omega_*$ limit in the power coefficients of $r$. In that case we can
ignore terms of order $(\tilde\omega_*^2, a_*^2, a_*\tilde\omega_*)$ or higher,
and obtain $-(n+1)\beta\simeq j$, $(n+1)(\beta+D_*-2)\simeq -(j +n+1)$,
and $\nu \simeq j+(n+1)/2$. We then demand the matching of the
corresponding multiplicative coefficients, which leads to a constraint for
the far-asymptotic integration constants $B_{1}$ and $B_2$, namely
\bea &~& \hspace*{-1.5cm}B \equiv \frac{B_1}{B_2} = -\frac{1}{\pi}
\left(\frac{2} {\tilde\omega
r_h\,(1+a_*^2)^\frac{1}{n+1}}\right)^{2j+n+1}
\sqrt{\tilde E_{j\ell m}+a^2\tilde\omega^2+\left(\frac{n+1}{2}\right)^2} \nonumber \\[2mm]
&& \hspace*{-0.6cm} \times\,\frac{\Gamma^2\left(\sqrt{\tilde
E_{j\ell m}+a^2\tilde\omega^2+
\left(\frac{n+1}{2}\right)^2}\right) \Gamma(\alpha+\beta + D_*
-1)\, \Gamma(\alpha+\beta)\, \Gamma(2-2\beta - D_*)}{\Gamma(2\beta
+ D_*-2)\, \Gamma(2+\alpha -\beta - D_*)\,\Gamma(1+\alpha-\beta)}
\,, \label{eq:Beq-bulk} \eea
that guarantees the existence of a smooth, analytic solution for the radial
part of the wavefunction for all $r$, valid in the low-energy and low-rotation
limit. We stress that, in order to achieve a higher level of accuracy in our
analysis, no expansion is performed in the arguments of the Gamma functions.
This method has been used in the literature before to derive analytic solutions
for brane \cite{CEKT2} and bulk \cite{CEKT4} massless scalar fields.
In both cases, the analytic results were shown to be in excellent agreement
with the exact numerical ones in the low-energy regime and quite often
at the intermediate-energy regime too.

In the presence of the mass term, though, there is one more constraint that
needs to be satisfied for the perfect match to take place. In the low-energy
and low-angular-momentum limit, the expression for the parameter $\beta$,
Eq. (\ref{beta-bulk}), becomes
\begin{equation}
\beta \simeq \frac{1}{2}\,\biggl[1-\frac{1}{(n+1)}\,\sqrt{(2j+n+1)^2
+4 m_\Phi^2 r_h^2}\,\biggr]\,.
\end{equation}
For $j \geq 0$ and $n \geq 1$, we thus need to satisfy $m_\Phi r_h <1$.
In order to derive some quantitative results, let us assume that $M_*=1$ TeV
and $M_{BH}=5$ TeV. If we ignore for a moment the angular momentum of the
black hole and use the mass-horizon radius relation for a higher-dimensional
Schwarzschild black hole, we find $r_h \simeq (4-2)\,10^{-4}$ fm for
$n=1-7$, respectively \cite{Kanti}. Then, the aforementioned constraint on
the mass of the bulk scalar field translates to
\begin{equation}
m_\Phi < (0.5-1)\,{\rm TeV}\,, \qquad {\rm for} \quad n=1-7\,.
\end{equation}
If we reinstate the angular momentum of the black hole, then the value of
the black-hole horizon, for the same mass, becomes smaller since
$r_h^{n+1}=\mu/(1+a_*^2)$; therefore, the upper bound on the mass of
the scalar field increases further and becomes easier to satisfy.

In order to define the absorption probability, we finally expand the far-field
solution (\ref{FF-bulk}) for $r\rightarrow \infty$, and obtain
\bea R_{FF}(r) &\simeq&
\frac{1}{r^\frac{n+2}{2}\sqrt{2\pi\tilde\omega}}\left[(B_1+iB_2)\,
e^{-i\,\left(\tilde\omega r - \frac{\pi}{2}\,\nu -
\frac{\pi}{4}\right)} + (B_1-iB_2)\,e^{i\,\left(\tilde\omega r -
\frac{\pi}{2}\,\nu -
\frac{\pi}{4}\right)} \right]\nonumber \\
&=& A_{in}^{(\infty)}\,\frac{e^{-i\tilde\omega
r}}{r^\frac{n+2}{2}} + A_{out}^{(\infty)}\,\frac{e^{i\tilde\omega
r}}{r^\frac{n+2}{2}}\,, \label{far-asym}\eea
which readily leads to
\beq \left|{\cal A}_{j \ell m}\right|^2 =
1-\left|\frac{A_{out}^{(\infty)}}{A_{in}^{(\infty)}}\right|^2 =
1-\left|\frac{B_1-iB_2}{B_1+iB_2}\right|^2 =
\frac{2i\left(B^*-B\right)}{B B^* + i\left(B^*-B\right)+1}\,.
\label{Absorption-def} \eeq

The above expression, in conjunction with Eq. (\ref{eq:Beq-bulk}), is our final
analytic result for the absorption probability for massive scalar fields
emitted in the bulk by a higher-dimensional, simply-rotating black hole.
Summarizing all of the aforementioned assumptions, it is valid as long
as the energy and mass of the emitted particle and the angular-momentum
of the black hole stay below unity (in units of $r_h^{-1}$ and $r_h$,
respectively). Its range of validity will be shortly investigated in
terms of the values of the above parameters, as well as that of the number
of extra dimensions $n$.

Equation (\ref{Absorption-def}) is also useful for studying analytically
various aspects of the absorption probability such as its behaviour in the
superradiant regime and the asymptotic limit $\tilde\omega \rightarrow 0$.
If we expand Eq. (\ref{Absorption-def}) in the low-energy limit, a more
convenient form may be derived for both purposes -- a similar analysis
was presented in all detail in \cite{CEKT2} where the emission of massless
scalar fields on the brane by the same type of black hole was
studied. From Eq. (\ref{eq:Beq-bulk}) we see that, in that limit, we obtain
$B \propto 1/\tilde\omega^{2j+n+1}$, and therefore
\beq \left|{\cal A}_{j\ell m}\right|^2 \simeq  2i\left(\frac{1}{B}
- \frac{1}{B^*}\right)=\Sigma_1 \times \Sigma_2 \times \Sigma_3\,,
\label{Abs-low-bulk} \eeq
where
\begin{equation}
\Sigma_1=\frac{-2i\pi\,\left(\tilde\omega r_h/2\right)^{2j+n+1}}
{(j+\frac{n+1}{2})\,\Gamma^2(j+\frac{n+1}{2})} \frac{(1+a_*^2)^{\frac{2j+n+1}{n+1}}\,
\Gamma(2\beta +D_*-2)}{\Gamma(2-2\beta -D_*)}\,,
\end{equation}
\begin{equation}
\Sigma_2=\frac{1}{|\Gamma(\alpha+\beta+D_*-1)|^2\,|\Gamma(\alpha+\beta)|^2}\,,
\label{Sigma2-bulk}
\end{equation}
and
\bea
\Sigma_3&=&\Gamma(2+\alpha-\beta-D_*)\,\Gamma(-\alpha+\beta+D_*-1)\,\Gamma(1+\alpha-\beta)\,
\Gamma(-\alpha+\beta)-(cc)\nonumber\\[2mm]
&=& \frac{-\pi^2\,\sin(2\pi\alpha)\,\sin \pi(2\beta+D_*)}
{\sin\pi(\alpha+\beta+D_*)\,\sin\pi(-\alpha+\beta+D_*)\,
\sin\pi(\alpha+\beta)\,\sin\pi(-\alpha+\beta)}\,. \label{Sigma3-bulk}
\eea
The $(cc)$ term above stands for the complex conjugate of the corresponding
expression. As the energy of the emitted mode decreases, moving towards the
asymptotic limit $\tilde \omega \rightarrow 0$, for modes with $m>0$, we
meet the value $\omega = \omega_s \equiv m \Omega_h$. From Eqs. (\ref{beta-bulk})
and (\ref{k}), it is clear that for that value $\alpha \rightarrow 0$, in which
case Eq. (\ref{Abs-low-bulk}) gets simplified to
\beq
\left|{\cal A}_{j\ell m}\right|^2 =\frac{4\pi \left(\tilde\omega r_h/2\right)^{2j+n+1} K_*
\,\sin^2\pi(2\beta+D_*)\,\Gamma^2(2\beta +D_*-2)\,\Gamma^2(1-\beta)\,(2-D_*-2\beta)}
{A_*\,(1+a_*^2)^{-\frac{2j +n+1}{n+1}}\,(j+\frac{n+1}{2})\,\Gamma^2(j+\frac{n+1}{2})\,
\Gamma^2(\beta+D_*-1)\,\sin^2\pi(\beta+D_*)}\,. \label{abs-super}
\eeq
In the above expression, all terms are positive definite, including the
$(2-D_*-2\beta)$ one, apart from $K_*$ whose sign, as expected, defines
the sign of the absorption probability: for $\omega < \omega_s$,
$(\omega-m\,\Omega_h)$ takes a negative value signalling the occurence of
superradiance.

For modes with $m \leq 0$, there is no superradiance effect, and we
may thus approach the asymptotic limit $\tilde \omega \rightarrow 0$.
From the coefficient $(\tilde \omega r_h)^{2j+n+1}$ in the expression of
$\Sigma_1$ it is clear that, in the massive case, too, it is the lowest
partial modes that dominate the value of the absorption probability in the
low-energy regime. We will therefore focus our attention on the dominant
mode $j=\ell=m=0$, and derive the behaviour of the absorption probability
in the above asymptotic limit. Although for massive modes with $m \leq 0$
the parameter $\alpha$ never becomes exactly zero, it acquires its smallest
possible value as $\tilde \omega \rightarrow 0$. Equation (\ref{abs-super})
therefore remains approximately valid, and, for $j=\ell=m=0$ and
$\beta=0 + {\cal O}(\tilde \omega^2)$, it is simplified
further to give
\beq \left|{\cal A}_{000}\right|^2 = \frac{4\pi
(1+a_*^2)^2(\tilde\omega r_h)^{n+1} \omega r_h} {A_* \,
2^n(n+1)\Gamma^2\left(\frac{n+1}{2}\right)(2-D_*)} + \ldots \,\,.
\eeq
We may also compute the absorption cross-section $\sigma_{000}$ for the dominant
massive scalar bulk mode in the asymptotic low-energy regime by using the
formula \cite{cross-section, CEKT4}
\beq \sigma_{j\ell m}(\omega) = \frac{2^n}{\pi}
\Gamma^2\left(\frac{n+3}{2}\right) \frac{A_H}{(\tilde\omega
r_h)^{n+2}}\,\frac{N_\ell}{(1+a_*^2)}\left|{\cal A}_{j\ell
m}\right|^2 \,, \label{cross-section} \eeq
that relates the absorption cross-section with the absorption probability for
a scalar mode propagating in the background of a higher-dimensional, simply
rotating black hole. In the above
\beq N_\ell = \frac{(2\ell+n-1)(\ell+n-2)!}{\ell!\,(n-1)!}\,, \qquad A_H =
\frac{
2\pi^{\frac{n+3}{2}}r_h^n\,(r_h^2+a^2)}{\Gamma\left(\frac{n+3}{2}\right)}
\label{Nell} \eeq
are the multiplicity of the $\ell$-th partial wave in the expansion of
the wave function over the hyperspherical harmonics on the
$n$-sphere \cite{CEKT4}, and the horizon area of the $(4+n)$-dimensional
rotating black hole, respectively. Substituting for the absorption
coefficient, we obtain
\beq \sigma_{000}(\omega) \simeq  \frac{(n+1)(1+a_*^2)
A_H}{A_*(2-D_*)} \left(\frac{\omega}{\tilde\omega}\right) +
\ldots\,. \eeq
For $a_*= 0$ and $m_\Phi=0$, the above reduces to the horizon area $A_H$
of a higher-dimensional, spherically-symmetric black hole, as was found
in \cite{HK1}. For $a_* \neq 0$ and $m_\Phi=0$, it was shown in
\cite{CEKT4,Jung-rot}
that the value of $\sigma_{000}$ remains very close to the area of the
corresponding rotating black hole as long as $a_*$ is not large. For
$m_\Phi \neq 0$, we observe significant deviations from this behaviour
as the value of the absorption cross-section for the lowest partial mode
is not only energy-dependent but deviates as $\tilde \omega \rightarrow 0$
-- this is in accordance with previous results derived in the cases of a
massive field propagating in the background of a 4-dimensional Kerr
\cite{Page} or of a $(4+n)$-dimensional, spherically-symmetric black
hole \cite{Jung-rot}. This behaviour is observed only in the case of the
lowest mode; higher modes have a $\tilde\omega^{2j+n+1}$ leading factor in
their absorption probability, and a $\tilde\omega^{2j-1}$ dependence for
their absorption cross-section -- for $j\geq 1$, this leads to a vanishing
value in the asymptotic limit $\tilde\omega \rightarrow 0$.

\begin{figure}[t]
  \begin{center}
  \mbox{\includegraphics[width = 0.51 \textwidth] {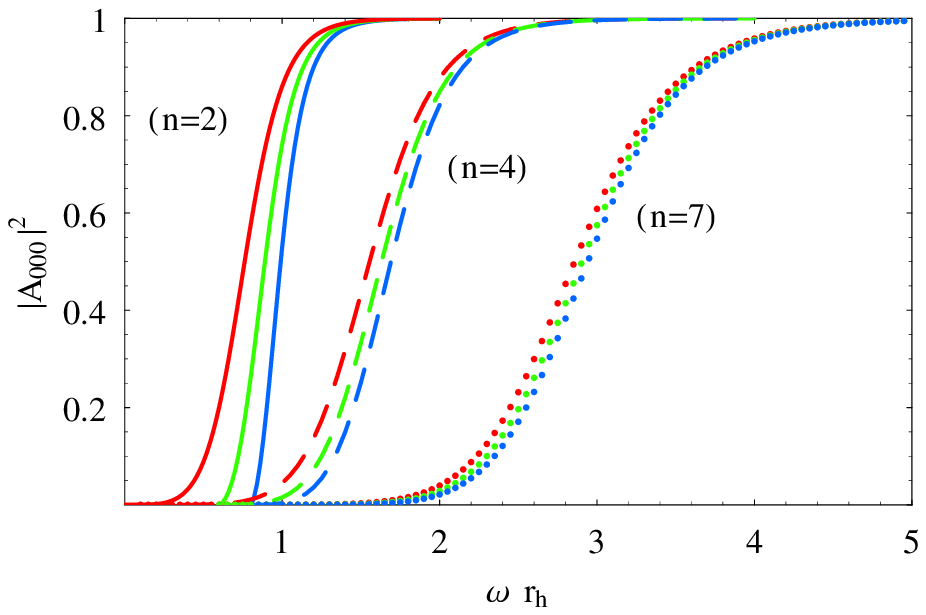}}
\hspace*{-0.8cm} {\includegraphics[width = 0.51 \textwidth]
{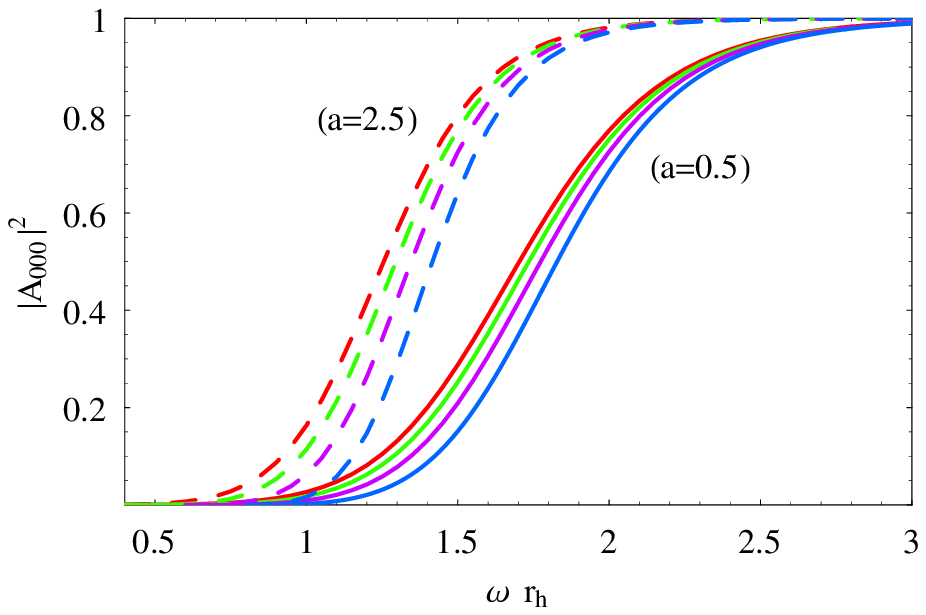}}
    \caption{ Absorption probabilities for the bulk scalar mode $j=\ell=m=0$, for
\textbf{(a)} $a_*=1$, $m_\Phi=0,0.6, 0.8$ (from left to right) and various $n$,
 and \textbf{(b)} $n=4$, $m_\Phi=0,0.4, 0.6, 0.8$ (from left to right) and various $a_*$.}
   \label{anmcombo}
  \end{center}
\end{figure}

For the derivation of the value of the absorption probability, that would be
valid for arbitrary values of the energy of the emitted particle and angular
momentum of the black hole, we need to solve Eq. (\ref{eq:radial-bulk})
numerically. To this end, a {\tt MATHEMATICA} code was constructed that
numerically solved for the value of the radial function $R(r)$ from the
horizon outwards. The boundary conditions for the second order differential
equation was the value of $R(r)$ and its first derivative at the horizon.
The asymptotic solution (\ref{near-asym}) was used for that purpose, with
the boundary conditions at $r \rightarrow r_h$ having the form
\begin{equation}
R=1\,, \qquad \frac{dR}{dr}=-ik\, \frac{dy}{dr}=
-\frac{ik (1+a_*^2)}{\Delta(r)}\,.
\label{conditions}
\end{equation}
The first condition was imposed to ensure that $|A_{-}|^2=1$ since no outgoing
mode is allowed to exist at the horizon. The second follows readily
from the asymptotic solution (\ref{near-asym}) and the use of the first
condition. The integration proceeds until we reach radial infinity (in
practice, this happens for $r \simeq 1000 r_h$) where, according to
Eq. (\ref{far-asym}), the radial function is a superposition of incoming
and outgoing modes. The corresponding amplitudes are then isolated and
the value of the absorption coefficient follows by use of the definition
(\ref{Absorption-def}).

As a consistency check, we have succesfully reproduced the numerical results
presented in \cite{CDKW2} for the case of massless scalar fields emitted
in the bulk by a simply rotating black hole - the case with $m_\Phi=0$
is also included in our plots for the easy comparison with the massless case.
In Fig. \ref{anmcombo} we plot the absorption probability for the
dominant mode $j=\ell=m=0$ as a function of the three parameters, $m_\Phi$,
$n$ and $a$, respectively. Figure \ref{anmcombo}(a) was drawn for fixed
angular-momentum parameter ($a_*=1$), and depicts the dependence of
$|{\cal A}_{000}|^2$ on the value of mass of the field and number of extra dimensions:
we observe that as $m_\Phi$ increases the value of the absorption
probability decreases as expected, since a larger amount of energy is
necessary for the emission of an increasingly more massive field. This
pattern holds independently of the value of $n$, nevertheless, the
suppression with $m_\Phi$ becomes less important as the number of extra
dimensions gets larger. Figure \ref{anmcombo}(a) reveals also that the
suppression of the absorption probability with the number of extra dimensions,
found previously for massless scalar fields in the bulk \cite{CEKT4, CDKW2},
holds also for massive fields. In Fig. \ref{anmcombo}(b), we keep fixed
the number of extra dimensions ($n=4$) and vary $m_\Phi$ and $a_*$: again
the suppression with the mass of the field is evident - contrary to what
happens with $n$, the suppression is more prominent as $a_*$ increases,
particularly in the low- and intermediate-energy regimes. The enhancement
of the absorption probability as $a_*$ itself increases, found again
previously in \cite{CEKT4, CDKW2}, persists also in the massive case.

\begin{figure}[t]
  \begin{center}
  \mbox{\includegraphics[width = 0.33 \textwidth] {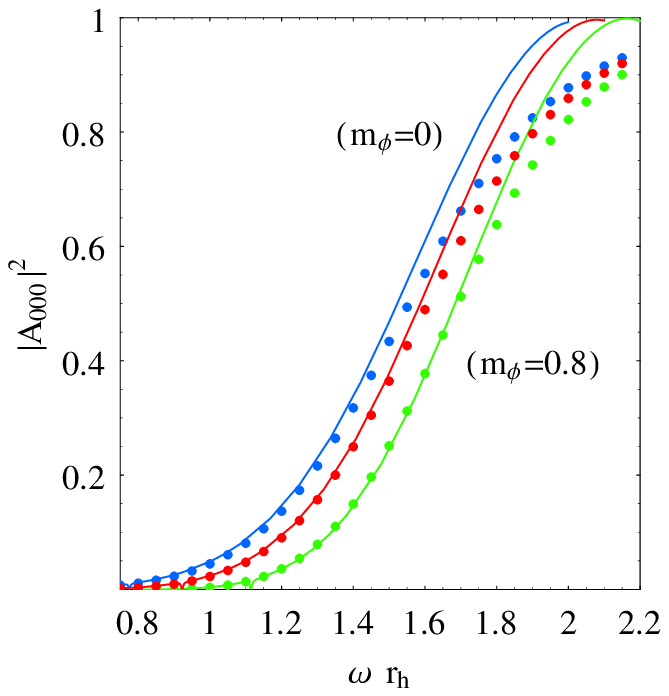}}
\hspace*{-0.4cm} {\includegraphics[width = 0.33 \textwidth]
{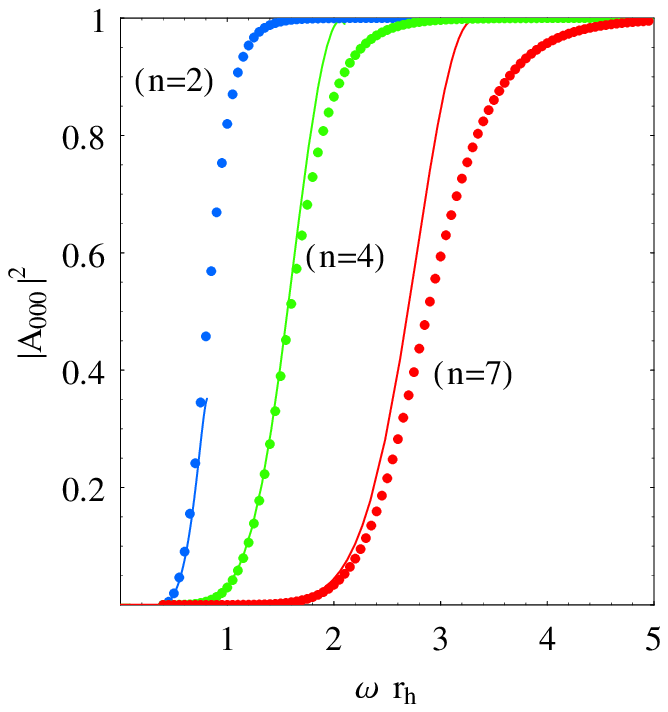}}
\hspace*{-0.3cm} {\includegraphics[width = 0.33 \textwidth]
{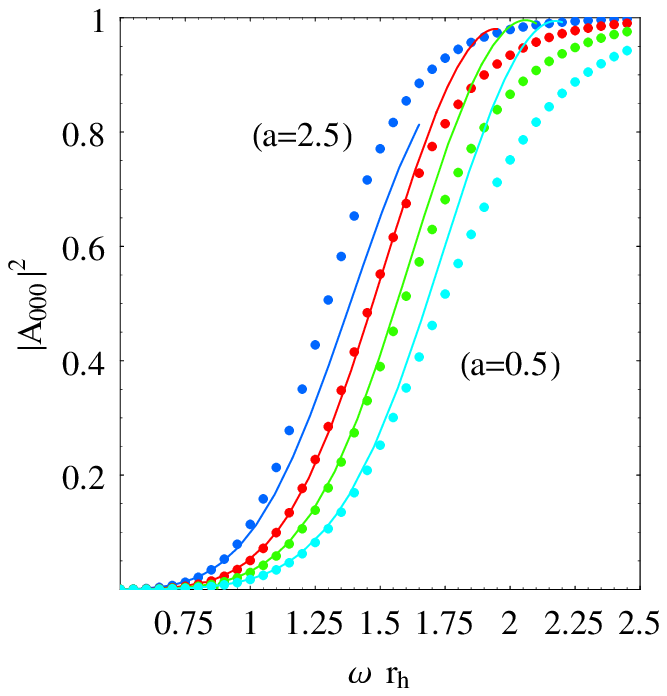}}
    \caption{ Comparison of the analytical (solid lines) and exact numerical
	(data points) results for the absorption probability for the bulk scalar
	mode $j=\ell=m=0$, for \textbf{(a)} $a_*=1$, $n=4$ and $m_\Phi=0,0.5,0.8$,
	\textbf{(b)} $a_*=1$, $m_\Phi=0.4$ and $n=2,4,7$, and  \textbf{(c)}
	$n=4$, $m_\Phi=0.4$ and $a_*=0.5,1,1.5,2.5$.}
   \label{anmcomp}
  \end{center}
\end{figure}

It would be interesting to compare the exact numerical results for the
value of the absorption probability with the ones following from the
analytical expression (\ref{Absorption-def}) with $B$ given by
Eq. (\ref{eq:Beq-bulk}). In Fig. \ref{anmcomp} we plot both sets of results
for a range of values of the parameters $m_\Phi$, $n$ and $a_*$ -- we
consider again the dominant scalar bulk mode $j=\ell=m=0$. Figure
\ref{anmcomp}(a) reveals that the agreement between numerical and
analytical results holds for a wide range of values of the mass parameter
below unity (in units of $r_h^{-1})$, as indeed expected from the
discussion below Eq. (\ref{eq:Beq-bulk}) regarding
the values of $m_\Phi$. On the other hand, in terms of the number of
extra dimensions, the agreement is case-dependent: as we see from
Fig. \ref{anmcomp}(b), it is remarkably good for $n=4$, for $n=7$ it is
limited in the lower part of the curves while for $n=2$ it stops abruptly
as the analytical result suffers from the existence of poles in the
arguments of the Gamma functions that force the value of $|{\cal A}_{000}|^2$
to dip towards smaller values and eventually vanish. The expression for
$B$, Eq. (\ref{eq:Beq-bulk}), is clearly the result of an approximation method
valid for small values of the angular-momentum parameter, and thus we
expect the agreement between the two sets of results to become worse
as the value of $a_*$ increases gradually; however, in Fig. \ref{anmcomp}(c),
we see that the agreement is actually improving as the angular-momentum
parameter increases reaching values even beyond unity, a result that
holds only in the presence of the mass term of the scalar field.

\begin{figure}[t]
  \begin{center}
  \mbox{\includegraphics[width=0.33 \textwidth] {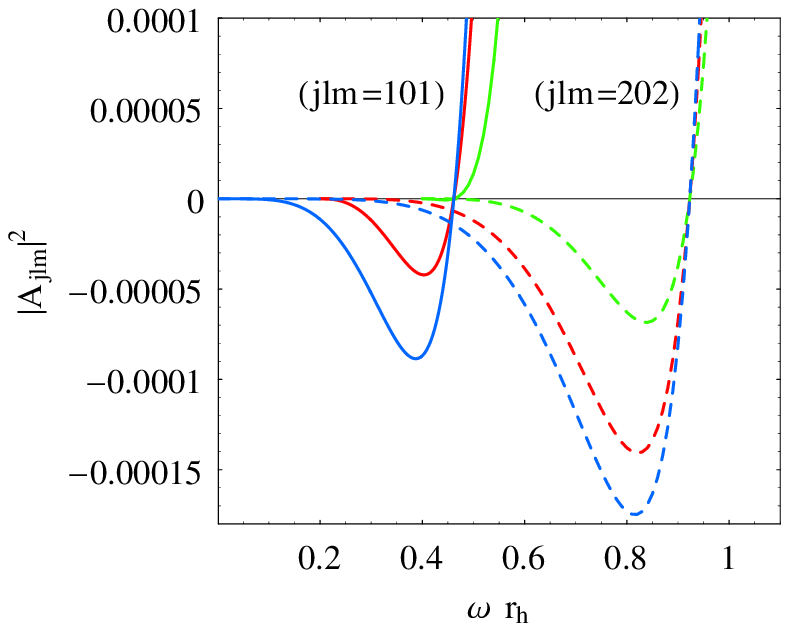}}
\hspace*{-0.3cm} {\includegraphics[width = 0.33 \textwidth]
{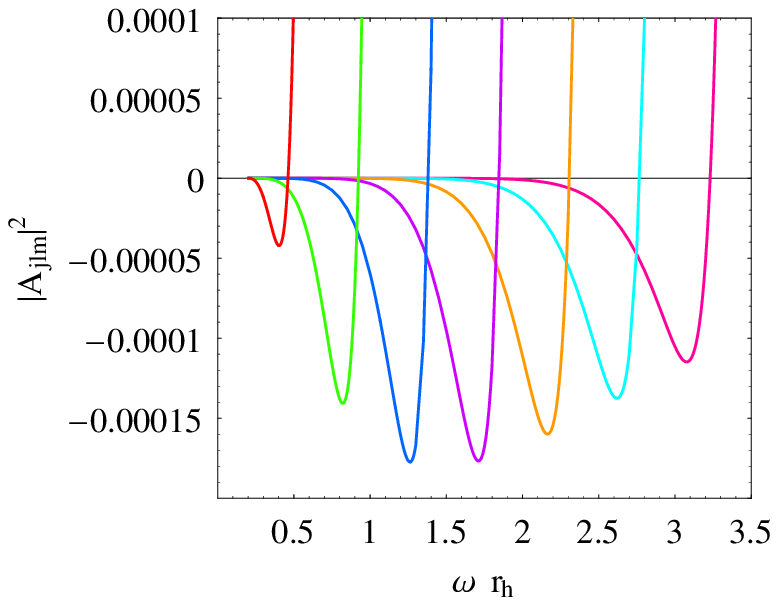}}
\hspace*{-0.3cm} {\includegraphics[width = 0.33 \textwidth]
{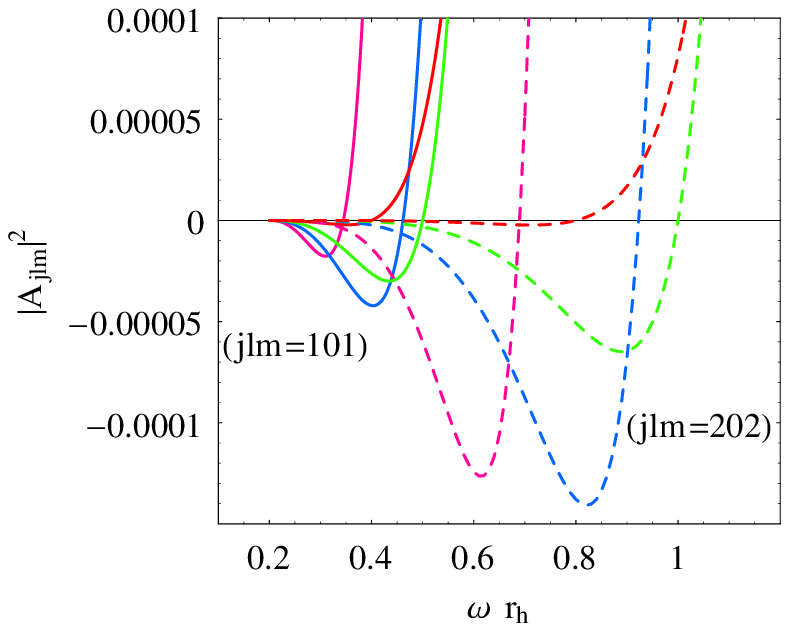}}
    \caption{ The dependence of superradiance for bulk massive scalar modes
	on \textbf{(a)} the mass, for $a_*=1.5$, $n=2$ and $m_\Phi=0,0.2,0.4$ (from bottom
	to top), \textbf{(b)} on the angular momentum numbers, for $a_*=1.5$, $n=2$, $m_\Phi=0.2$
	and $j=m=1,2,3,4,5,6,7$ (from left to right), and  \textbf{(c)} on the angular
	momentum parameter, for $n=2$, $m_\Phi=0.2$ and $a_*=0.5\,(\rm red),1\,(\rm green),
	1.5\,(\rm blue), 2.5\,(\rm magenta)$.}
   \label{super-bulk}
  \end{center}
\end{figure}

Let us finally comment on the behaviour of the superradiance \cite{super} on
the parameters of the theory. In the context of the general suppression of
the value of the absorption probability as the mass of the field increases,
we expect that the effect of the superradiance will also be suppressed --
this is indeed depicted in Fig. \ref{super-bulk}(a) where the value
of $|{\cal A}_{j\ell m}|^2$ is plotted for various values of $m_\Phi$ for
the indicative modes $(j\ell m=101)$ and $(j\ell m=202)$. Despite the observed
dominance of the superradiance effect for the mode $(j\ell m=202)$ over the
one for $(j \ell m=101)$, this pattern does not hold indefinitely as the angular
momentum numbers increase: in fact, from Fig. \ref{super-bulk}(b), where we plot the
superradiant regime for the modes $j=m=1,2,...,7$ for $m_\Phi=0.2$, $n=2$
and $a_*=1.5$, it is clear that the mode $j=m=3$ is the dominant superradiant
one, a result that was also found in the massless case \cite{CDKW2}.
The suppression of the superradiance with the number of extra
dimensions observed in \cite{CDKW2} for massless bulk scalar modes
holds also in the massive case, and thus we do not comment further.
A feature that has not been noted before is the non-monotonic behaviour
of both the magnitude of the superradiance effect and the extent of the
superradiant regime in terms of the angular-momentum parameter $a_*$: in
Fig. \ref{super-bulk}(c), we see that, as $a_*$ increases from zero to 1.5,
the superradiance effect is indeed enhanced, however, this behaviour
is reversed when $a_*$ increases further. In addition, superradiance
occurs for frequencies $m_\Phi < \omega < \omega_s=ma/(r_h^2+a^2)$: the
latter restriction is imposed by the vanishing of the value of the absorption
probability; the former by the demand that its value is a real number,
and signifies the fact that no
particles of mass $m_\Phi$ can be created if energy less than that is
available. Interestingly enough the width of the superradiance regime,
$\delta \omega =ma/(r_h^2+a^2)-m_\Phi$, does not monotonically grow with the
increase of the angular momentum of the black hole, as one could instictively
expect. Indeed, its value reaches a maximum for a particular value of
the angular-momentum parameter, namely $a=\pm r_h$, which is in fact
independent of the mass and angular momentum numbers of the mode as well
as of the number of extra dimensions \footnote{The monotonic behaviour
of the width and depth of the superradiance regime found in \cite{CEKT4}
is not in contradiction with the results found here as only low values of the
angular momentum of the black hole, lower than the turning points found
here and in agreement with the low-$a_*$ approximation used in
\cite{CEKT4}, were considered in there.}. For the case depicted in
Fig. \ref{super-bulk}(c), where we have fixed the horizon value at $r_h=1$
and considered only positive values of $a_*$,
the superradiant regime takes its maximum value at $a_*=1$, beyond which
it starts to shrink, for both modes $(j\ell m=101)$ and $(j\ell m=202)$.


\subsection{Energy Emission Rate in the Bulk}

We will next compute the rate of energy emission in the bulk in the form
of massive scalar fields by using the exact numerical results for the
absorption probability found in the previous section. The emission of
energy per unit frequency and unit time in the bulk is given by the
expression \cite{Page, CEKT4, CDKW2}
\beq \frac{d^2 E}{dt d\omega} = \frac{1}{2\pi}\sum_{j, \ell,
m}\frac{\omega}{\exp\left[k/T_\text{H}\right]-1}\,N_\ell \left|{\cal
A}_{j\ell m}\right|^2 \,. \label{rate-bulk}\eeq
The multiplicity of states $N_\ell$ from the expansion of the wavefunction
of the field in the $n$-dimensional sphere is given in Eq. (\ref{Nell}) and
the parameter $k$ is defined in Eq. (\ref{k}). Finally, the temperature of
the higher-dimensional, simply-rotating black hole (\ref{rot-metric-bulk}) is
\begin{equation}
T_\text{H}=\frac{(n+1)+(n-1)a_*^2}{4\pi(1+a_*^2)r_{h}}\,.
\label{temp}
\end{equation}
Equation (\ref{rate-bulk}) is identical in form with the expression for
the emission of massless scalar fields in the bulk, nevertheless, there
are two major differences: the calculation of the spectrum starts from
$\omega=m_\Phi$ instead of zero, and the value of the absorption
probability depends, apart from the spacetime parameters, on the
characteristics of the emitted field including its mass.

\begin{figure}[t]
  \begin{center}
\mbox{\includegraphics[width = 0.45 \textwidth] {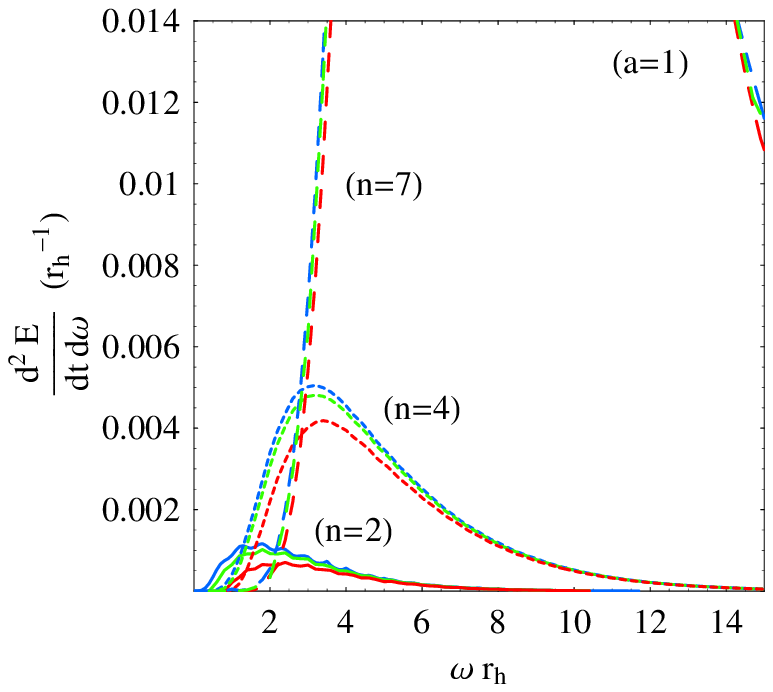}}
\hspace*{0.2cm} {\includegraphics[width = 0.45 \textwidth]
{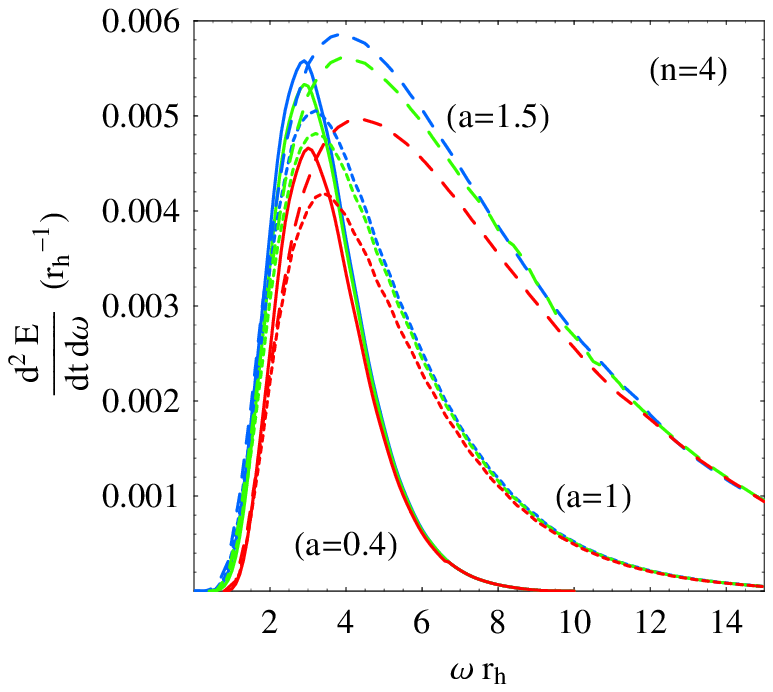}}
    \caption{ Energy emission rates for bulk scalar fields for: \textbf{(a)}
	for $a_*=1$, $n=2,4,7$ and $m_\Phi=0,0.4,0.8$ (from top to bottom in each
	set of curves with fixed $n$), and \textbf{(b)} $n=4$, $a_*=0.4,1,1.5$ and
	$m_\Phi=0,0.4,0.8$ (from top to bottom in each set of curves with fixed $a_*$).}
   \label{power-bulk}
  \end{center}
\end{figure}


In order to derive the energy emission spectrum, we need to sum over a
significantly large number of partial waves labeled by the set of ($j,\ell,m$)
quantum numbers. For each value of $n$, $a_*$ and $m_\Phi$, we aimed at
deriving the complete spectrum, i.e. to reach values of the energy parameter
$\omega r_h$ where the corresponding value of the energy emission rate would
be less than $10^{-6}$. At the same time, the number of partial waves summed
had to be large enough so that the  derivation of the energy spectrum would
be as close as possible to the real one -- especially for the computation of
the total emissivity presented in section 4. Taking all these constraints
into account, we were able to sum the contribution of all bulk scalar modes
up to $j=30$, that brings the total number of summed modes to
$N_{\rm bu}=5456$. According to our estimates, the contribution of all modes
higher than $j=30$ should be less than $5\%$, for the higher values
of parameters considered, namely $n=7$ and $a_*=1.5$, an error that falls
below 0.001$\%$ for the lowest values considered, i.e. $n=2$ and $a_*=0.4$.

In Fig. \ref{power-bulk}, we depict the energy emission rate on the brane
in the form of massive scalar fields in terms of the number of extra dimensions,
value of the angular-momentum parameter, and mass of the emitted field itself.
Thus, Fig. \ref{power-bulk}(a) shows the energy emission rate for fixed $a_*$
($a_*=1$) and variable $n=2,4,7$ and $m_\Phi=0,0.4,0.8$, while
Fig.\ref{power-bulk}(b) plots the same quantity but for fixed $n$ ($n=4$)
and variable $a_*=0.4,1,1.5$. In terms of the spacetime parameters $n$
and $a_*$, these plots confirm the behaviour found in the case of massless
fields \cite{CEKT4, CDKW2}: the power spectrum is enhanced as the number of
extra dimensions increases while its dependence on the angular momentum
parameter is not monotonic but differs as $n$ and/or $\omega$ varies.
More detailed features, like the oscillatory pattern of the emission curves
for low values of $n$ and $a_*$, that are replaced by more smoother curves
as the values of these parameters increase, are also recovered.

In terms of the mass of the scalar field, we observe the expected suppression
of the emission rate, for fixed $n$ and $a_*$, as $m_\Phi$ increases --
the suppression is more prominent in the low- and intermediate-energy
regimes whereas the effect of the mass becomes negligible at the
high-energy regime. Compared to the case of the emission of masless
scalar fields, the suppression in the low-energy regime becomes even more
significant if the disappareance of the frequency range with $\omega < m_\Phi$
is taken into account. The magnitude of the suppression with $m_\Phi$ depends
strongly on the particular value of $n$ and $a_*$ -- the exact effect will
be computed in section 4 where the total emissivities in bulk and brane
will be calculated.


\section{Emission of Massive Scalars on the Brane}

In this section, we turn our attention to the emission of massive scalar
fields by a higher-dimensional simply-rotating black hole on the brane.
The analysis for the derivation of the absoprtion probability, both
analytical and numerical, is quite similar to the one performed for the
emission in the bulk; aspects of it have also been recently addressed in
a set of publications \cite{Sampaio} that appeared while this
work was still in progress. For the sake of comparison and completeness
of the analysis, we will still present in this section the most important
points of our calculation on the brane and focus our discussion to aspects not
covered before; these include, for example, the analytic study of the low-energy
asymptotic behaviour of the absorption probability and cross-section,
the role of the angular momentum of the black hole, that was ignored in
\cite{Sampaio}, and the form of the energy emission spectrum, instead
of the number flux that was studied in the same work.

Let us start with the form of the gravitational background that a massive
scalar field sees as it propagates on the brane and its corresponding field
equation. The 4-dimensional induced background will be the projection of
the higher-dimensional one (\ref{rot-metric-bulk}) onto the brane, and
follows by fixing the values of the angular variables of the $n$-sphere.
Then, the induced-on-the-brane line-element takes the form
\begin{equation}
\begin{split}
ds^2=\left(1-\frac{\mu}{\Sigma\,r^{n-1}}\right)dt^2&+\frac{2 a\mu\sin^2\theta}
{\Sigma\,r^{n-1}}\,dt\,d\varphi-\frac{\Sigma}{\Delta}dr^2 \\[3mm] &\hspace*{-1cm}
-\Sigma\,d\theta^2-\left(r^2+a^2+\frac{a^2\mu\sin^2\theta}{\Sigma\,r^{n-1}}\right)
\sin^2\theta\,d\varphi^2\,,
\end{split} \label{induced}
\end{equation}
which is very similar to the usual 4-dimensional Kerr one but carries
an explicit dependence on the number of additional spacelike dimensions
$n$. The field equation is still given by the covariant form
(\ref{field-eq-bulk}) but with the higher-dimensional metric tensor
$G_{MN}$ replaced by the 4-dimensional one $g_{\mu\nu}$ defined above.
The field factorization
\begin{equation}
\Phi(t,r,\theta,\varphi)= e^{-i\omega t}\,e^{i m \varphi}\,P(r)\,T(\theta)\,,
\end{equation}
leads again to the decoupling of variables and to the following set of
radial and angular equations
\begin{equation}
\frac{d}{dr}\biggl(\Delta\,\frac{d P}{dr}\biggr)+
\left(\frac{K^2}{\Delta} -\tilde\Lambda_{jm}-m_\Phi^2 r^2
\right)P=0\,, \label{radial-brane}
\end{equation}
\smallskip
\begin{equation}
\label{angular} \frac{1}{\sin\theta}
\frac{d}{d\theta}\left(\sin\theta\,\frac{d T} {d\theta}\right) +
\biggl(\tilde\omega^2a^2\cos^2\theta -\frac{m^2}{\sin^2\theta}
+ \tilde E_{jm}\biggr)T=0\,,
\end{equation}
respectively. In the above, we have defined
$\tilde\Lambda_{jm}=\tilde E_{jm}+a^2\omega^2-2am\omega$, while
$\tilde \omega$ is again given by $\tilde\omega=\sqrt{\omega^2-m_\Phi^2}$
and $K$  by Eq. (\ref{eq:K}). The angular function $T(\theta)$ satisfies again
a modified spheroidal harmonics equation with $\omega \rightarrow \tilde\omega$.
The corresponding massive eigenvalue $\tilde E_{jm} (a\tilde\omega)$
is thus related to the massless one through the same shift, and in terms
of a power series \cite{eigenvalue} is given by
\bea
\tilde E_{jm} &=& j\,(j+1)+ (a\tilde\omega)^2\,\frac{[2m^2-2j\,(j+1)+1]}{(2j-1)\,(2j+3)}
\nonumber \\[1mm] &+&
(a\tilde\omega)^4\,\left\{\frac{2\,[-3+17 j\,(j+1) +j^2 (j+1)^2(2j-3)\,(2j+5)]}
{(2j-3)\,(2j+5)\,(2j+3)^3(2j-1)^3} \right. \nonumber\\[1mm]
&+& \left. \frac{4m^2}{(2j-1)^2(2j+3)^2}\,
\left[\frac{1}{(2j-1)\,(2j+3)} -\frac{3j\,(j+1)}{(2j-3)\,(2j+5)}\right] \right.
\nonumber \\[1mm] &+& \left.
\frac{2m^4\,[48 +5 (2j-1)\,(2j+3)]}{(2j-3)\,(2j+5)\,(2j-1)^3(2j+3)^3}\right\} +
{\cal O}\Bigl((a \tilde \omega)^6\Bigr)\,,
\eea
The above form will be used in the computation of the absorption probability
both analytically and numerically.

\subsection{The Absorption Probability on the Brane}

The approximation method employed in section 2 can again be used to solve
the radial equation (\ref{radial-brane}) analytically. The same change
of variable $r \rightarrow f(r) = \Delta(r)/(r^2+a^2)$, in the near-horizon
regime ($ r \simeq r_h$), leads to an equation of the form (\ref{eq:NH-1-bulk})
where now
\begin{equation}
D_* \equiv 1+ \frac{n\,(1+a_*^2)}{A_*} - \frac{4 a_*^2}{A_*^2}\,, \qquad
C_* \equiv (\tilde{\Lambda}_{jm} + m_\Phi^2r_{h}^2)\,(1+a_*^2)\,,
\end{equation}
while $A_*$ and $K_*$ are defined as in the bulk. The field redefinition
$P(f)=f^\alpha (1-f)^\beta F(f)$ reduces the above differential equation
to a hypergeometric one with the physically acceptable solution in the
near-horizon regime given by
\beq
P_{NH}(f)=A_- f^{\alpha}\,(1-f)^\beta\,F(a,b,c;f)\,. \label{NH-brane}
\eeq
In the above, $A_-$ is again an arbitrary integration constant, and
$a=\alpha + \beta +D_*-1$, $b=\alpha + \beta$, $c=1 + 2 \alpha$.
The power coefficients $\alpha$ and $\beta$ are given by the expressions
in Eq. (\ref{beta-bulk}), with $D_*$ and $C_*$ now taken their brane values.
Under the choice $\alpha=\alpha_-$, that we will henceforth use, the above
solution reduces, as expected, to an ingoing plane wave,
$P_{NH} \simeq A_-\,f^{-iK_*/A_*}=A_-\,e^{-iky}$ with $k$ defined in
Eq. (\ref{k}).

In the far-field regime ($r \gg r_h$), the radial equation (\ref{radial-brane}),
under the substitution $P(r)= \frac{1}{\sqrt{r}}\,\tilde P(r)$, takes again
the form of a Bessel differential equation leading to the general solution
\beq
P_{FF}(r)=\frac{B_1}{\sqrt{r}}\,J_{\nu}\,(\tilde\omega r) +
\frac{B_2}{\sqrt{r}}\,Y_{\nu}\,(\tilde\omega r) \,,
\label{FF-brane}
\eeq
where now $\nu=\sqrt{\tilde E_{jm}+a^2\tilde\omega^2+1/4}$.

The process of the matching proceeds as in the case of bulk emission.
The near-horizon solution (\ref{NH-brane}), after it is shifted, is expanded
in the limit $r \gg r_h$, while the far-field one (\ref{FF-brane}) is expanded
in the $r \rightarrow 0$ limit. Both reduce to polynomial forms similar
to those in Eqs. (\ref{NH-stretched}) and (\ref{FF-stretched}).  If we
again ignore terms of order $(\tilde\omega_*^2, a_*^2, a_*\tilde\omega_*)$
or higher in the power coefficients, we obtain $-(n+1)\beta\simeq j$,
$(n+1)(\beta+D_*-2)\simeq -(j+1)$, and $\nu \simeq j+1/2$. These simplifications
hold provided that the mass of the scalar field on the brane does not exceed
an upper value: following a similar argument to the case of the bulk
emission, this constraint is found to be $m_\Phi < (250-500)\,{\rm GeV}$
for $n=1-7$ -- note that the upper value of the mass on the brane is reduced
by a factor of two compared to the one in the bulk. Then, the matching of the
corresponding multiplicative in the coefficients, leads to the constraint
\bea
&~& \hspace*{-1.8cm}B \equiv \frac{B_1}{B_2} = -\frac{1}{\pi}
\left(\frac{2}{\tilde\omega r_h\,(1+a_*^2)^\frac{1}{n+1}}\right)^{2j+1}
\sqrt{\tilde E_{jm}+a^2\tilde\omega^2+1/4} \nonumber \\[2mm]
\hspace*{0.2cm}
&& \times\,\frac{\Gamma^2\left(\sqrt{\tilde E_{jm}+a^2\tilde\omega^2+1/4}
\right)\Gamma(\alpha+\beta + D_* -1)\,\Gamma(\alpha+\beta)\,
\Gamma(2-2\beta - D_*)}{\Gamma(2\beta + D_*-2)\,
\Gamma(2+\alpha -\beta - D_*)\,\Gamma(1+\alpha-\beta)} \,. \label{Beq-brane}
\eea
The above completes the derivation of the analytic solution for the radial
part of the massive scalar field on the brane. By expanding the far-field
solution (\ref{FF-brane}) at asymptotic infinity, we recover again a
superposition of spherical waves
\bea
P_{FF}(r) &\simeq& \frac{1}{\sqrt{2\pi\tilde\omega}}\left[\frac{(B_1+iB_2)}{r}\,
e^{-i\,\left(\tilde\omega r - \frac{\pi}{2}\,\nu - \frac{\pi}{4}\right)}
+ \frac{(B_1-iB_2)}{r}\,e^{i\,\left(\tilde\omega r - \frac{\pi}{2}\,\nu -
\frac{\pi}{4}\right)} \right]\,. \label{FF-infinity-brane}
\eea
The absorption probability for the brane emission $|{\cal A}_{jm}|^2$ is then
given again by the right-hand-part of Eq. (\ref{Absorption-def}) with $B$ in
this case defined in Eq. (\ref{Beq-brane}).

At the very low-energy regime, we may again derive a simplified, compact
expression for the absorption probability. Following the same analysis as
in the case of bulk emission, we obtain
$|{\cal A}_{jm}|^2=\Sigma_1 \times \Sigma_2 \times \Sigma_3$, where
\begin{equation}
\Sigma_1=\frac{-2i\pi\,\left(\tilde\omega r_h/2\right)^{2j+1}}
{(j+\frac{1}{2})\,\Gamma^2(j+\frac{1}{2})} \frac{(1+a_*^2)^{\frac{2j+1}{n+1}}\,
\Gamma(2\beta +D_*-2)}{\Gamma(2-2\beta -D_*)}\,,
\end{equation}
while $\Sigma_2$ and $\Sigma_3$ are given by the corresponding bulk equations
(\ref{Sigma2-bulk}) and (\ref{Sigma3-bulk}) but with the parameters $\alpha$,
$\beta$ and $D_*$ now taken their brane values. For modes with $m>0$, the
limit $\alpha \rightarrow 0$, will give us the behaviour of the absorption
probability at the upper boundary of the superradiance regime which is given by
\beq
\left|{\cal A}_{jm}\right|^2 =\frac{4\pi \left(\tilde\omega r_h/2\right)^{2j+1} K_*
\,\sin^2\pi(2\beta+D_*)\,\Gamma^2(2\beta +D_*-2)\,\Gamma^2(1-\beta)\,(2-D_*-2\beta)}
{A_*\,(1+a_*^2)^{-\frac{2j+1}{n+1}}\,(j+\frac{1}{2})\,\Gamma^2(j+\frac{1}{2})\,
\Gamma^2(\beta+D_*-1)\,\sin^2\pi(\beta+D_*)}\,. \label{abs-super-brane}
\eeq
As expected it is again the sign of $K_*$ that defines the sign of the
absorption probability in this energy regime since $K_*= r_h (1+a_*^2)(\omega -
m \Omega_h)$. By setting $j=m=0$ and expanding further in the limit
$\omega \rightarrow 0$, Eq. (\ref{abs-super-brane}) can also give us the
asymptotic value of $|{\cal A}_{00}|^2$ for the dominant scalar mode,
which is
\beq
\left|{\cal A}_{\,00}\right|^2 = \frac{4\,\omega \tilde\omega
r_h^2\,(1+a_*^2)} {A_*\,(1+a_*^2)^{-1/(n+1)}\,(2-D_*)} + ... \,.
\eeq
Equation (\ref{cross-section}) may also provide the relation between the
absorption cross-section and the absoprtion probability for a massive scalar
field living on the brane. By setting $n=0$ and $N_\ell=1$, since the brane
modes do not `see' the $n$-sphere, we obtain the 4-dimensional formula
\beq \sigma_{\,00} = \frac{\pi}{\tilde\omega^2}\,\left|{\cal
A}_{\,0}\right|^2= 4\pi\
\left(\frac{\omega}{\tilde\omega}\right)(r_h^2+a^2)\,\frac{(1+a_*^2)^{1/(n+1)}}
{\left[(n+1)+(n-1)\,a_*^2\right]\,(2-D_*)}+ ... \,.
\label{lowsigma} \eeq
Again, for $m_\Phi=0$ and $a_*=0$, the value of the absorption cross-section
reduces to the area $4\pi r_h^2$ of the 4-dimensional Schwarzschild black
hole, as expected \cite{KMR, HK1}; for $a_* \neq 0$, it approaches the area
$4 \pi (r_h^2+a^2)$ of the 4-dimensional rotating black hole for small values
of the angular-momentum parameter \cite{CEKT2}. However, as soon as the
mass of the scalar field becomes larger than zero, the aforementioned constant
values  of $\sigma_{\,00}$ are replaced by diverging ones for both rotating
and non-rotating black holes -- in the latter case, this is again in accordance
with previous analyses \cite{Page, Jung-rot}. As in the case of the bulk scalar
field, and due to the $\tilde \omega ^{2j+1}$ factor in Eq. (\ref{abs-super-brane}),
all higher modes with $j \geq 1$ have a vanishing asymptotic value as
$\tilde \omega \rightarrow 0$.

The derivation of the complete energy spectrum demands once again the
calculation of the value of the absorption probability by numerical means.
The asymptotic behaviour of the brane massive scalar field close to and far
away from the black hole horizon is similar to the one of a bulk field: it
is an incoming plane wave in the near-horizon regime, as discussed below
Eq. (\ref{NH-brane}), and a spherical wave at radial infinity according
to Eq. (\ref{FF-brane}). The numerical integration of the
radial differential equation (\ref{radial-brane}) on the brane is
performed by using the same method as in the bulk: the integration starts
very close to the black-hole horizon with boundary conditions given again by
Eq. (\ref{conditions}) and proceeds until we reach radial infinity, where
the amplitudes of the incoming and outgoing modes are isolated to
compute the value of the absorption probability $|{\cal A}_{jm}|^2$.

\begin{figure}[t]
  \begin{center}
  \mbox{\includegraphics[width = 0.51 \textwidth] {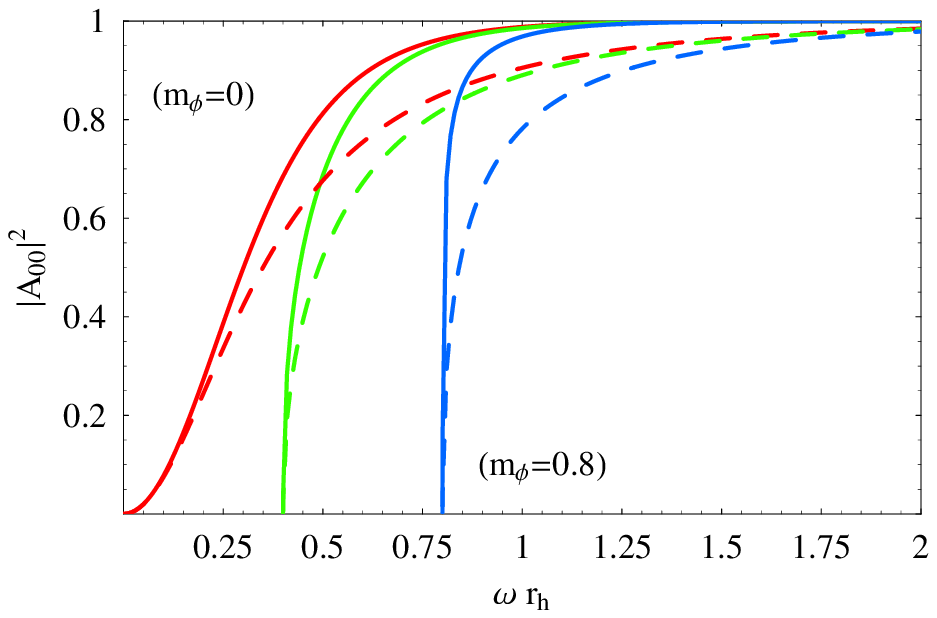}}
\hspace*{-0.8cm} {\includegraphics[width = 0.51 \textwidth]
{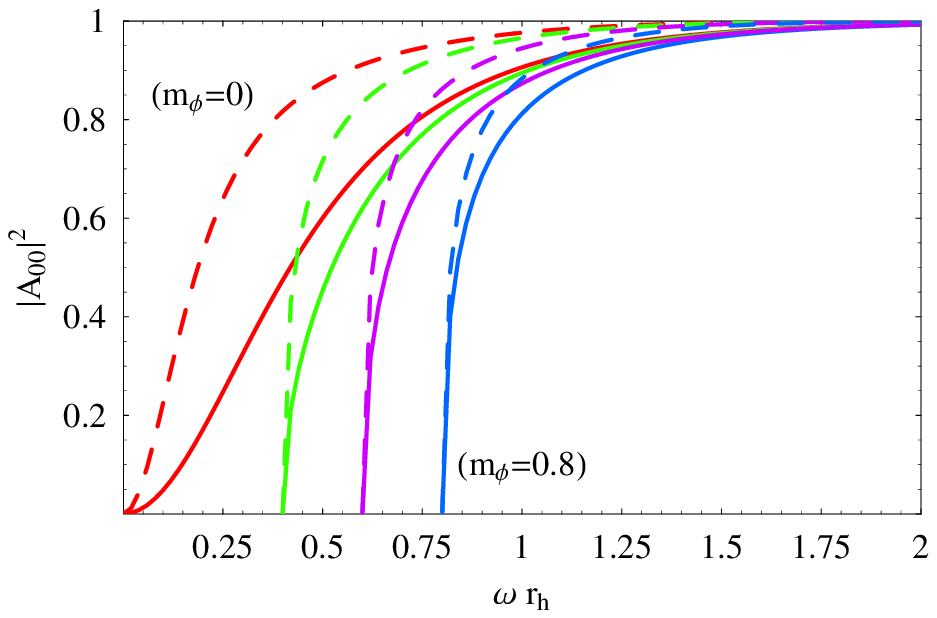}}
    \caption{ Absorption probabilities for the brane scalar mode $j=m=0$, for
\textbf{(a)} $a_*=1$, $m_\Phi=0,0.4, 0.8$ and $n=2$ (\rm solid lines) and
7 (\rm dashed lines), and \textbf{(b)} $n=4$, $m_\Phi=0,0.4, 0.6, 0.8$ and $a_*=0.5$
 (solid lines) and 2.5 (dashed lines).}
   \label{grey-brane-n-mphi}
  \end{center}
\end{figure}

Therefore, in Fig. \ref{grey-brane-n-mphi}, we depict the exact numerical
results for the value of the absorption probability of the dominant mode
$j=m=0$. In Fig. \ref{grey-brane-n-mphi}(a), the value of $|{\cal A}_{00}|^2$
is plotted for fixed angular momentum ($a_*=1$) and variable $m_\Phi$,
equal to 0, 0.4, 0.8 (from left to right), and for two different values
of the number of extra dimensions, $n=2$ (solid lines) and $n=7$ (dashed lines).
As expected, the value of the absorption probability is suppressed with
the number of extra dimensions, as noted before in the literature \cite{HK2,
DHKW, CEKT2}. The suppression becomes significantly more important as the
value of the mass of the brane scalar field increases -- this is also in
agreement \footnote{The agreement is mainly qualitative as our results
are derived for non-vanishing angular momentum parameter $a_*$ while in
\cite{Sampaio} the effect of the rotation of the black hole was ignored
and the role of the mass and charge of the brane field was studied instead.
Nevertheless, there is a general agreement between the two sets of results
in terms of both the number of extra dimensions and the value of the mass
of the brane scalar field.} with the results derived recently in
\cite{Sampaio}, therefore, we do not comment on this further.
On the other hand, Fig. \ref{grey-brane-n-mphi}(b) depicts the dependence
of $|{\cal A}_{00}|^2$ on the angular-momentum parameter, that takes the
values $a_*=0.5$ and $a_*=2.5$, while $n$ remains fixed ($n=4$) and
$m_\Phi$ changes from 0 to 0.8 (from left to right again). For $m_\Phi=0$,
the absorption probability increases as $a_*$ increases, too, in accordance
again with the literature \cite{HK2, DHKW, CEKT2} - the same behaviour is
observed as the mass of the scalar field becomes larger but with the
enhancement becoming increasingly less significant. For the purpose of
the analysis presented in section 4, where the bulk and brane energy spectra
are compared, let us note here that both effects, the suppression with $n$
and the enhancement with $a_*$, are much more prominent for massive bulk
scalar fields than for brane fields of the same type.

\begin{figure}[t]
  \begin{center}
\mbox{\includegraphics[width = 0.4 \textwidth] {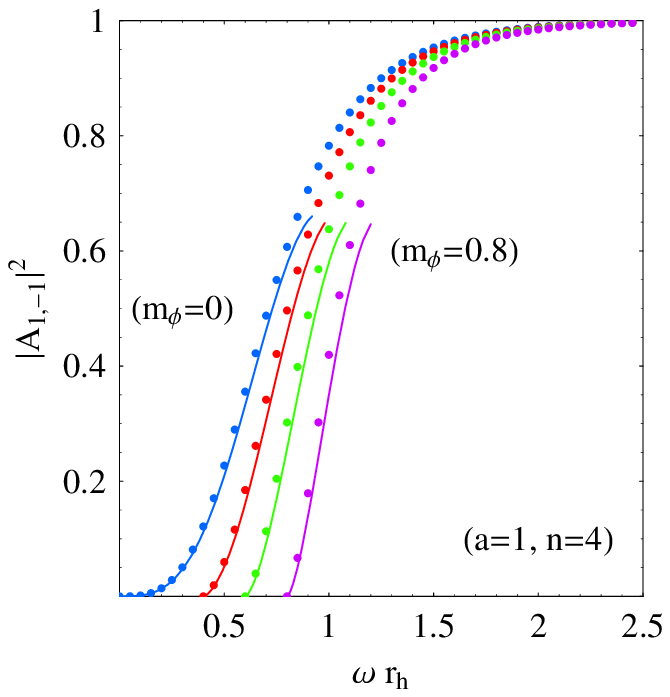}}
\hspace*{0.5cm} {\includegraphics[width = 0.4 \textwidth]
{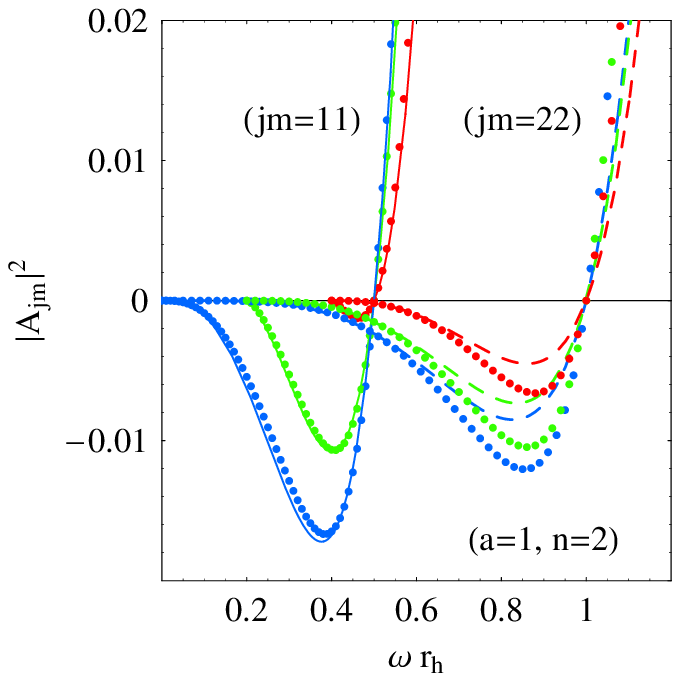}}
    \caption{ Comparison of the analytical (solid lines) and exact numerical
	(data points) results for the absorption probability for: \textbf{(a)}
	the brane scalar mode $(j,m)=(1,-1)$, for $a_*=1$, $n=4$ and $m_\Phi=0,0.4,0.6,0.8$,
	and \textbf{(b)} the superradiant brane scalar modes $(j,m)=((1,1),(2,2))$ for
	$a_*=1$, $n=2$, and  $m_\Phi=0,0.2,0.4$ (from bottom to top).}
   \label{comp-brane-mphi}
  \end{center}
\end{figure}

As in the case of the emission in the bulk, we would like to investigate
the validity of the analytic method used above to derive the value of
the absorption probability for the emission of massive scalar fields
on the brane, and how this is affected by the value of the mass and
angular-momentum numbers of the emitted field, the number of extra
dimensions and the magnitude of the angular momentum of the black hole.
To this end, in Fig. \ref{comp-brane-mphi}(a), we plot both the analytical
(solid lines) and numerical (data points) results for the absorption
probability of the indicative mode $(j,m)=(1,-1)$, for fixed angular-momentum
parameter ($a_*=1$) and number of extra dimensions ($n=4$) and
$m_\Phi=0,0.4,0.6,0.8$. We observe that the agreement between the
two sets of results remains particularly good even well beyond the
low-energy regime. The solid lines terminate again due to the existence
of poles in the arguments of the Gamma functions in the analytic
expression of the absorption coefficient. We find that the appearance of
the poles, for modes with given $j$, takes place much earlier for the
modes with $m=0$ than for $m >0$, while for the ones with $m < 0$ this
happens at much higher values of the energy, a fact which significantly
extends the range of validity of the analytic results in the latter case
as is clear from Fig. \ref{comp-brane-mphi}(a). In Fig. \ref{comp-brane-mphi}(b),
we focus on the low-energy regime of two superradiant modes, $(j,m)=(1,1)$
and $(j,m)=(2,2)$: in agreement with results drawn in the massless case
\cite{CEKT2}, we find that the analytic results for the value of the
absorption probability start to deviate from the exact numerical ones as
the angular-momentum numbers of the mode increase; this is due to the
shift of the curve towards higher values of the energy -- note that
the range of agreement extends well beyond the value $\omega r_h=0.6$
for both modes, however, for the $(j,m)=(2,2)$ mode this covers only a
part of the superradiant regime contrary to what happens for the
$(j,m)=(1,1)$ mode. It deserves to be noted that the value of the
mass of the emitted field affects the relative values of the absorption
probability of different superradiant modes: while for $m_\Phi=0$, the $(j,m)=(1,1)$
mode dominates over the $(j,m)=(2,2)$ one, this radically changes as
soon as the mass of the scalar field exceeds the value $m_\Phi=0.2$.

\begin{figure}[t]
  \begin{center}
\mbox{\includegraphics[width = 0.4 \textwidth] {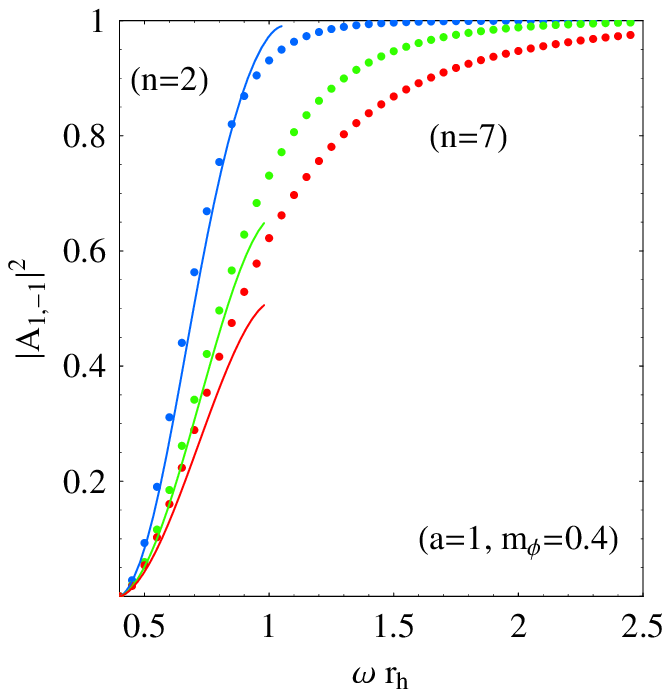}}
\mbox{\includegraphics[width = 0.4 \textwidth] {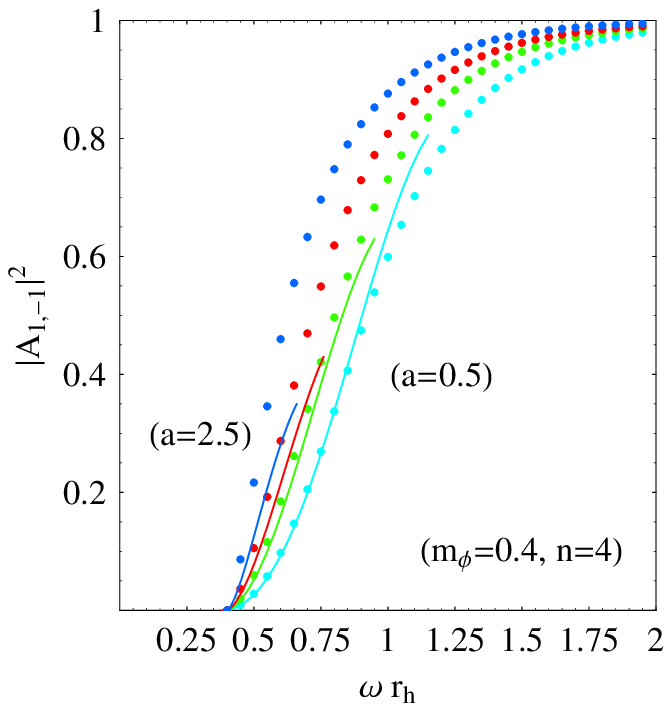}}
    \caption{ Comparison of the analytical (solid lines) and exact numerical
	(data points) results for the absorption probability for the brane scalar
	mode $(j,m=1,-1)$, for \textbf{(a)} $a_*=1$, $m_\Phi=0.4$ and $n=2,4,7$,
	and  \textbf{(b)} $m_\Phi=0.4$, $n=4$ and $a_*=0.5,1,1.5,2.5$.}
   \label{ancomp-brane}
  \end{center}
\end{figure}

Let us finally comment on the range of validity of the analytic
results in terms of the parameters of the higher-dimensional spacetime.
In Fig. \ref{ancomp-brane}(a), we plot both sets of results for the
mode $(j,m=1,-1)$ for fixed angular-momentum parameter ($a_*=1$) and
mass of the field ($m_\Phi=0.4$) while the number of extra dimensions
takes the values $n=2,4,7$. From the plot, it is clear that the
agreement between the analytic and numerical results is excellent
for low values of $n$ while it quickly worsens as the number of extra
dimensions increases. Figure \ref{ancomp-brane}(b) plots the two
sets of results for the same mode for fixed mass ($m_\Phi=0.4$) and
number of extra dimensions ($n=4$), but variable angular-momentum
parameter ($a_*=0.5,1,1.5,2.5$). Again, the agreement extends well
beyond the intermediatate-energy regime for low values of $a_*$ while
it is gradually restricted in the low-energy regime as the value of
$a_*$ increases. The observed behaviour is in agreement with the one
found in the massless case \cite{CEKT2} and stems from the fact that
several of our approximations in the analytic method become less
accurate as either $n$ or $a_*$ increases.


\subsection{Energy Emission Rate on the Brane}

The exact value of the absorption probability $|{\cal A}_{jm}|^2$ for massive
scalar fields on the brane, as this followed after the numerical integration
of the radial equation of the wavefunction, will now be used for the computation
of the corresponding energy emission rate. The higher-dimensional, simply-rotating
black hole emits massive scalar particles on the brane with a rate given by
the exression \cite{IOP, HK2, IOP2, DHKW}
\beq \frac{d^2 E}{dt d\omega} = \frac{1}{2\pi}\sum_{j,m}
\frac{\omega}{\exp\left[k/T_\text{H}\right]-1}\, \left|{\cal
A}_{jm}\right|^2 \,.
\label{rate-brane}\eeq
In the above, $k$ is defined in Eq. (\ref{k}) as before, while the temperature
for the emission on the brane is that of the higher-dimensional black hole
given in Eq. (\ref{temp}). As in the case of bulk emission, the formula of
the emission rate for massive fields is the same as the one for massless,
with the effect of the mass being encoded in the value of the absoprtion
probability and the frequency range of the emission.

\begin{figure}[t]
  \begin{center}
\mbox{\includegraphics[width = 0.44 \textwidth] {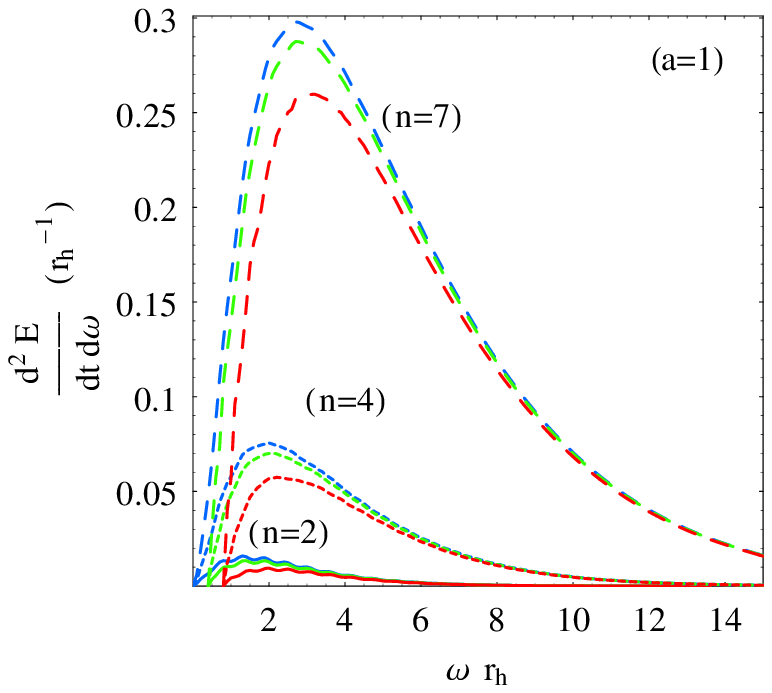}}
\hspace*{0.2cm} {\includegraphics[width = 0.44 \textwidth]
{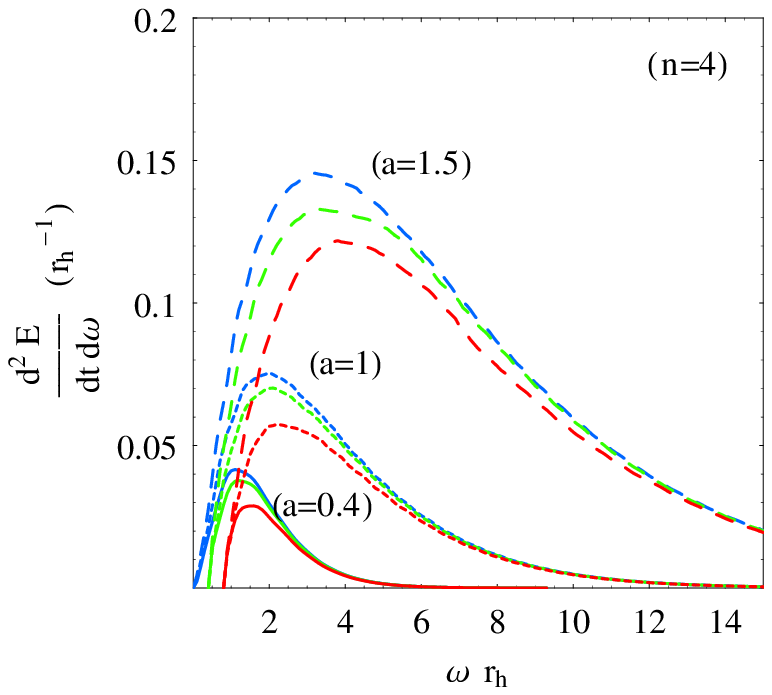}}
    \caption{ Energy emission rates for brane scalar fields for: \textbf{(a)}
	for $a_*=1$, $n=2,4,7$ and $m_\Phi=0,0.4,0.8$ (from top to bottom in each
	set of curves with fixed $n$), and \textbf{(b)} $n=4$, $a_*=0.4,1,1.5$ and
	$m_\Phi=0,0.4,0.8$ (from top to bottom in each set of curves with fixed $a_*$).}
   \label{power-brane}
  \end{center}
\end{figure}

As in the case of the bulk emission, for the derivation of the energy emission
spectrum on the brane we need to sum over a significantly large number of partial waves
labeled by the ($j,m$) quantum numbers. The absence of the `internal' quantum
number $\ell$, that further characterizes the bulk modes, makes the brane
summation easier, nevertheless the process remained significantly time-consuming
\footnote{For the largest values of the parameters considered, i.e. $n=7$,
and $a_*=1.5$, the derivation of the complete spectrum for each value of
the mass $m_\Phi$ lasted more than 4 days - this is to be contrasted with
the corresponding summation in the bulk where a single run lasted more than
6 days.}. We summed the contribution of all modes up to $j=40$, that brings
the total number of brane modes to $N_{br}=1681$, and computed the spectrum up
to the value of energy where the power rate dropped again below $10^{-6}$.
According to our estimates, the error in our results by leaving out the higher
modes is less than $5\%$, for the higher values of $n$ and $a_*$ considered,
and below 0.001$\%$ for the lowest.

In Fig. \ref{power-brane}, we plot the energy emission rate on the brane
in the form of massive scalar fields in terms of the number of extra dimensions,
value of the angular-momentum parameter, and mass of the emitted field
-- we have kept the same values of these parameters as in the case of
bulk emission for easier comparison. Figure \ref{power-brane}(a) shows
the energy emission rate for fixed $a_*$ ($a_*=1$) and variable $n=2,4,7$
and $m_\Phi=0,0.4,0.8$, while in Fig. \ref{power-brane}(b) we keep $n$
fixed ($n=4$) and vary $a_*=0.4,1,1.5$. Again, our results
reproduce succesfully the behaviour found in the case of massless fields
\cite{HK2,DHKW,CEKT4}, and demonstrate that the enhancement of the emission
spectrum as either the number of extra dimensions or the angular momentum
of the black hole increases persists even for non-vanishing values of
the mass of the emitted field. The mass of the scalar field causes again the
suppression of the spectrum in all energy regimes, apart from the
very high-energy one where its effect becomes negligible. The suppression
is again strongly dependent on the particular value of $n$ and $a_*$.
By comparing Figs. \ref{power-bulk} and \ref{power-brane}, we see that
the brane emission is larger than the bulk emission by more than an order
of magnitude - accordingly, we expect the suppression with $m_\Phi$ to
be larger on the brane than in the bulk. The exact role of $m_\Phi$ in
the total emissivity of the black hole, in conjunction with the parameters
($n$, $a_*$) and the type of emission channel (brane or bulk), will be
investigated in the next section.


\section{Bulk and Brane Total Emissivities}

Although the global properties of the absorption probability and energy
spectra do not change when the mass of the scalar field is introduced,
important variations in their values appear which differ as the value
of $m_\Phi$, together with that of either $n$ or $a_*$, changes. For
this reason, we expect that differences will appear when the spectra for
the emission of massive and massless fields are compared. These differences
may be evident at particular energy regimes or range of values of the
parameters $(n,a_*)$, and may significantly affect the total energy
emissivities. The modifications in the spectrum may also be different
when bulk or brane emission is considered, therefore, in this section
we compute the total emissivities for both emission channels and
compare them.

To this end, we have integrated the differential energy rates per unit
time and unit frequency, computed in sections 2 and 3, over the entire frequency
range of emission. In Tables 1 and 2, we present the corresponding total
emissivities for bulk and brane emission, respectively, for some indicative
values of the number of extra dimensions ($n=2,4,7$), angular-momentum
parameter of the black hole ($a_*=0.4,1,1.5$) and mass of the emitted
field ($m_\Phi=0,0.4,0.8$). The values of the total emissivities are
normalised to the one for $n=2$, $a_*=0.4$ and $m_\Phi=0$, in each
case, for easy comparison.

The entries of both tables confirm the enhancement of the total emissivities
as either $n$ or $a_*$ increases and the suppression with $m_\Phi$.
As it was anticipated from the plots, the suppression is strongly dependent
not only on $m_\Phi$ but also on both $n$ and $a_*$. We obsereve that,
as either $n$ or $a_*$ increases, the suppression of the
total emissivity with the mass of the scalar field decreases in magnitude.
Starting from the bulk channel (Table 1), we see that for a fixed, low
value of $n$, i.e. $n=2$ the total emission for a scalar field with
mass $m_\Phi=0.8$ drops to 52$\%$ of the emission for a massless
field, if $a_*=0.4$, but to 71$\%$ if $a_*=1.5$. The suppression is
even more limited when the value of $n$ takes a much higher value: thus,
for $n=7$, the emission for a $m_\Phi=0.8$ scalar field in the bulk drops
only to the 92\% of the massless value if $a_*=0.4$ and to 95$\%$ if
$a_*=1.5$. It seems that both the number of extra dimensions and the
rotation of the black hole subsidize the emission of massive scalar fields.

\begin{table}[t]
\begin{center}
\caption{Total energy emissivities for massive scalar fields in the bulk}
\bigskip
\begin{tabular}{
cl|rrr}
\hline
\hspace*{0.2cm}  &  \hspace*{0.2cm} & \hspace*{0.4cm} $a_*=0.4$ &
\hspace*{0.4cm} $a_*=1.0$  & \hspace*{0.4cm} $a_*=1.5$ \\[1mm] \hline
 $n=2$ \hspace*{0.2cm}& \hspace*{0.2cm} $m_\Phi=0$ \hspace*{0.1cm} &
1.00 \hspace*{0.2cm} & 1.54	\hspace*{0.2cm} &	3.46 \hspace*{0.2cm} \\
 & \hspace*{0.2cm} $m_\Phi=0.4$ \hspace*{0.2cm} & 0.84	 \hspace*{0.2cm}
 &	1.34 \hspace*{0.2cm} &	3.05 \hspace*{0.2cm} \\
 & \hspace*{0.2cm} $m_\Phi=0.8$ \hspace*{0.2cm} & 0.52	 \hspace*{0.2cm}
 & 0.95	\hspace*{0.2cm} & 2.46 \hspace*{0.2cm} \\ \hline
$n=4$ \hspace*{0.2cm} & \hspace*{0.2cm} $m_\Phi=0$   & 6.29 \hspace*{0.2cm}
& 9.57	\hspace*{0.2cm} & 19.22 \hspace*{0.2cm} \\
& \hspace*{0.2cm} $m_\Phi=0.4$ \hspace*{0.2cm} & 5.97 \hspace*{0.2cm}
& 9.13  \hspace*{0.2cm}  & 18.61 \hspace*{0.2cm} \\
& \hspace*{0.2cm} $m_\Phi=0.8$ \hspace*{0.2cm} & 5.12 \hspace*{0.2cm}
& 7.99 \hspace*{0.2cm}	& 16.74 \hspace*{0.2cm} \\
 \hline
$n=7$ \hspace*{0.2cm} & \hspace*{0.2cm} $m_\Phi=0$  \hspace*{0.1cm}  &
131.47 \hspace*{0.2cm} & 202.48	\hspace*{0.2cm} & 327.37 \hspace*{0.2cm} \\
& \hspace*{0.2cm} $m_\Phi=0.4$ \hspace*{0.2cm} & 128.56 \hspace*{0.2cm}
& 197.27 \hspace*{0.2cm}   & 322.87 \hspace*{0.2cm} \\
& \hspace*{0.2cm} $m_\Phi=0.8$ \hspace*{0.2cm} & 121.57 \hspace*{0.2cm}
& 188.58 \hspace*{0.2cm}   & 310.18 \hspace*{0.2cm} \\ \hline
\end{tabular}
\label{emiss-bulk}
\end{center}
\end{table}
\begin{table}[t]
\begin{center}
\caption{Total energy emissivities for massive scalar fields on the brane}
\bigskip
\begin{tabular}{cl|rrr}
\hline
\hspace*{0.2cm}  &  \hspace*{0.2cm} & \hspace*{0.4cm} $a_*=0.4$ &
\hspace*{0.4cm} $a_*=1.0$  & \hspace*{0.4cm} $a_*=1.5$ \\[1mm] \hline
 $n=2$ \hspace*{0.2cm}& \hspace*{0.2cm} $m_\Phi=0$ \hspace*{0.1cm} &
1.00 \hspace*{0.2cm} & 3.37	\hspace*{0.2cm} &	13.18 \hspace*{0.2cm} \\
 & \hspace*{0.2cm} $m_\Phi=0.4$ \hspace*{0.2cm} & 0.75	\hspace*{0.2cm}
 &	3.10 \hspace*{0.2cm} &	11.98 \hspace*{0.2cm} \\
 & \hspace*{0.2cm} $m_\Phi=0.8$ \hspace*{0.2cm} & 0.39	\hspace*{0.2cm}
 & 2.16	\hspace*{0.2cm} & 9.51 \hspace*{0.2cm} \\ \hline
$n=4$ \hspace*{0.2cm} & \hspace*{0.2cm} $m_\Phi=0$   & 6.56 \hspace*{0.2cm}
& 25.73	\hspace*{0.2cm} & 89.89 \hspace*{0.2cm} \\
& \hspace*{0.2cm} $m_\Phi=0.4$ \hspace*{0.2cm} & 5.73 \hspace*{0.2cm}
& 23.75  \hspace*{0.2cm}  & 84.18 \hspace*{0.2cm} \\
& \hspace*{0.2cm} $m_\Phi=0.8$ \hspace*{0.2cm} & 4.14 \hspace*{0.2cm}
& 19.51 \hspace*{0.2cm}	& 83.39 \hspace*{0.2cm} \\
 \hline
$n=7$ \hspace*{0.2cm} & \hspace*{0.2cm} $m_\Phi=0$  \hspace*{0.1cm}  &
36.75 \hspace*{0.2cm} & 144.53	\hspace*{0.2cm} & 483.83 \hspace*{0.2cm} \\
& \hspace*{0.2cm} $m_\Phi=0.4$ \hspace*{0.2cm} & 34.48 \hspace*{0.2cm}
& 138.86 \hspace*{0.2cm}   & 471.08 \hspace*{0.2cm} \\
& \hspace*{0.2cm} $m_\Phi=0.8$ \hspace*{0.2cm} & 29.28 \hspace*{0.2cm}
& 126.53 \hspace*{0.2cm}   & 440.77 \hspace*{0.2cm} \\ \hline
\end{tabular}
\label{emiss-brane}
\end{center}
\end{table}

The same behaviour is observed for emission on the brane (Table 2) although
here the suppression is larger: for $n=2$ the total emission for a brane
scalar field with mass $m_\Phi=0.8$ drops to 39$\%$ of the emission for a
massless field, if $a_*=0.4$, but to 72\% if $a_*=1.5$; for $n=7$, the
emission for a $m_\Phi=0.8$ scalar field on the brane drops to the 80\% of
the massless value if $a_*=0.4$ and to 91\% if $a_*=1.5$.

Since the suppression of the total emissivities between brane and bulk
emission differ, it is imperative to calculate the relative emissivities to
find out whether the mass of the emitted field changes the energy balance
in the bulk-brane channels. These, derived by dividing the actual values
of the bulk and brane emissivities for each set of values ($n, a_*, m_\Phi$),
are displayed in Table 3. Our results confirm and extend the ones of
\cite{CDKW2} where the emission of massless scalar fields was studied.
In there, it was found that the bulk emission channel was becoming
increasingly sub-dominant as the value of the rotation parameter
increased from $a_*=0$ to $a_*=1$ - here, we show that this behaviour
persists for higher values of the angular momentum parameter. Also,
we confirm that the bulk-over-brane ratio take its lower value for
an intermediate value of the number of extra dimensions (a result
that was found in the case of both rotating \cite{CDKW2} and non-rotating
\cite{HK1} black holes) but starts increasing again as $n>4$.

\begin{table}[t]
\begin{center}
\caption{Bulk-over-brane relative energy emissivities for massive scalar fields}
\bigskip
\begin{tabular}{cl|rrr}
\hline
\hspace*{0.2cm}  &  \hspace*{0.2cm} & \hspace*{0.4cm} $a_*=0.4$ &
\hspace*{0.4cm} $a_*=1.0$  & \hspace*{0.4cm} $a_*=1.5$ \\[1mm] \hline
 $n=2$ \hspace*{0.2cm}& \hspace*{0.2cm} $m_\Phi=0$ \hspace*{0.1cm} &
0.180 \hspace*{0.2cm} & 0.076	\hspace*{0.2cm} &	0.0451 \hspace*{0.2cm} \\
 & \hspace*{0.2cm} $m_\Phi=0.4$ \hspace*{0.2cm} & 0.202	 \hspace*{0.2cm}
 &	0.078 \hspace*{0.2cm} &	0.0458 \hspace*{0.2cm} \\
 & \hspace*{0.2cm} $m_\Phi=0.8$ \hspace*{0.2cm} & 0.24	 \hspace*{0.2cm}
 & 0.079	\hspace*{0.2cm} & 0.0466 \hspace*{0.2cm} \\ \hline
$n=4$ \hspace*{0.2cm} & \hspace*{0.2cm} $m_\Phi=0$   & 0.173 \hspace*{0.2cm}
& 0.067	\hspace*{0.2cm} & 0.038 \hspace*{0.2cm} \\
& \hspace*{0.2cm} $m_\Phi=0.4$ \hspace*{0.2cm} & 0.188 \hspace*{0.2cm}
& 0.069  \hspace*{0.2cm}  & 0.039 \hspace*{0.2cm} \\
& \hspace*{0.2cm} $m_\Phi=0.8$ \hspace*{0.2cm} & 0.223 \hspace*{0.2cm}
& 0.074 \hspace*{0.2cm}	& 0.040 \hspace*{0.2cm} \\
 \hline
$n=7$ \hspace*{0.2cm} & \hspace*{0.2cm} $m_\Phi=0$  \hspace*{0.1cm}  &
0.645 \hspace*{0.2cm} & 0.253	\hspace*{0.2cm} & 0.122 \hspace*{0.2cm} \\
& \hspace*{0.2cm} $m_\Phi=0.4$ \hspace*{0.2cm} & 0.673 \hspace*{0.2cm}
& 0.256 \hspace*{0.2cm}   & 0.124 \hspace*{0.2cm} \\
& \hspace*{0.2cm} $m_\Phi=0.8$ \hspace*{0.2cm} & 0.749 \hspace*{0.2cm}
& 0.269 \hspace*{0.2cm}   & 0.127 \hspace*{0.2cm} \\ \hline
\end{tabular}
\label{emiss-bulk}
\end{center}
\end{table}

Overall, it is clear that the brane channel remains the dominant one
over the bulk channel, during the emission of both massless and massive
fields. Nevertheless, we find that the presence of the mass gives a
considerable boost to the bulk-over-brane energy ratio, especially
for low values of the angular momentum parameter. The boost depends also on
the number of extra dimensions: for $n=2$, the mass of a $m_\Phi=0.8$
scalar field increases the bulk-over-brane energy ratio of a black hole
with $a_*=0.4$ by 33\%, while for $n=7$ the increase is 16\%. We thus
conclude that, when the effect of the mass of the emitted field is taken
into account, it is the fast-rotating black holes living in a spacetime
with a fairly large number of extra dimensions that lose the smallest part
of their energy into invisible bulk emission.

Let us finally note that the results presented in this work not only extend
previous analyses for massless fields, but also improve those, too. For
instance, our results for the total bulk emissivities when $m_\Phi=0$ agree
in the first or second decimal point (depending on the value of $n$ and
$a_*$) with those derived in \cite{CDKW2} - the agreement is reassuring as
a different numerical code was used. Small deviations between our results
may be due to the fact that, in the calculation of the total emissivities,
we have not imposed any cut-off on the frequency but instead tried to obtain
the complete spectrum by keeping a realistically large number of scalar
modes.


\section{Conclusions}

In this work, we have moved towards the direction of considering the
emission of realistic particle states by a higher-dimensional, simply
rotating black hole. We have studied the emission of massive scalar fields
both in the bulk and on the brane, and investigated the role that the
mass of the field plays in the corresponding energy spectra profiles and
in the bulk-over-brane energy ratio.

The emission of Hawking radiation in the bulk in the form of massive
scalar fields was studied first. The radial part of the field equation
was first solved analytically, and an expression for the absorption
probability was found that helped us investigate low-energy aspects
of the emission. Next, by using numerical analysis, the exact value of
the absorption probability was determined and its dependence on the
mass of the emitted field, in conjunction with the number of extra
dimensions and angular-momentum of the black hole, was studied. As
expected, the presence of the mass term caused the suppression of the
absorption probability as additional energy is required for the
emission of a massive field. Our numerical and analytical results
were directly compared, and found to be in excellent agreement in the
low and intermediate energy regimes for scalar fields with a mass
smaller than (0.5-1) TeV.

The exact numerical value of the absorption probability was subsequently
used to derive the differential emission rate per unit time and unit
frequency in the bulk. Particular care was taken so that a large enough
number of scalar modes ($N_{bu} \simeq 5500$) was summed up in our
computation of the energy spectra. The mass term caused the suppression
of the energy spectra in the low and intermediate-energy regimes,
compared to the massless case: for low values of $n$ and $a_*$ and
$m_\Phi=0.8$, the suppression is of the order of 50\%, while it becomes
smaller in magnitude as either $n$ or $a_*$ increases.

The same task was performed for the emission of massive scalar fields
on the brane. The value of the absorption probability was again found
both analytically and numerically, and it was shown that the two sets
of results are in very good agreement, in the lowest part of the spectrum,
up to masses of order (250-500) GeV. The exact profile of the energy
spectra on the brane was found next in terms of the parameters
($m_\Phi, n, a_*$), with the mass term causing again a significant
suppression in their value. The suppression was larger than the one
in the bulk decreasing the value of the energy emission rate to
approximately 40\% of that in the massless case, for low values of
$n$ and $a_*$ and for $m_\Phi=0.8$. As in the case of bulk emission,
a considerable number of modes  ($N_{br} \simeq 1700$) was summed up
in our calculation so that the computed spectra are as close as
possible to the real ones.

The role of the mass of the emitted field in the bulk-over-brane
energy ratio was also investigated. The total energy emissivities
of bulk and brane emission were derived and directly compared.
In agreement with previous analyses \cite{EHM, HK1, CDKW2} --
that we have generalised by considering a larger range of parameters
of both $n$ and $a_*$ --  we found that the bulk channel remains
sub-dominant to the brane one; nevertheless, the bulk-over-brane
ratio takes a considerable value especially for a large number of
extra dimensions and a slowly rotating black hole. We further found
that the presence of the mass of the emitted field increases the
percentage of energy which is spent by the black hole in the bulk.
For a small number of extra dimensions and a low value of the
angular-momentum of the black hole, the enhancement of the bulk
channel over the brane one can reach the value of 33\% if $m_\Phi=0.8$.

In conclusion, in this work we have performed a comprehensive study of
the emission of massive scalar fields by a higher-dimensional, simply
rotating black hole both in the bulk and on the brane. We have studied
the dependence of the absorption probabilities and energy emission rates
on all parameters of the theory, and compared analytic and numerical
methods for the computation of their value. We have confirmed the
importance of the emission of a higher-dimensional black hole both in
the bulk and on the brane, and demostrated that properties of the
emitted field, such as its mass which was up to now largely ignored,
can play a significant role in the bulk-over-brane energy balance.

\bigskip

{\bf Acknowledgments.} We are deeply grateful to Athanasios Dedes for
useful discussions regarding our numerical code. We also acknowledge
participation in the RTN Universenet (MRTN-CT-2006035863-1 and
MRTN-CT-2004-503369).


\end{document}